\documentclass[reprint,amsmath,amssymb,aps,pra,groupedaddress,showpacs]{revtex4-1}
\usepackage{graphicx}%
\usepackage{dcolumn}%
\usepackage{bm}%

\usepackage[
   mag=1000,
  left=2.5cm, right=1cm, top=2cm, bottom=2cm, headsep=0.7cm, footskip=1cm
]{geometry}
\def\e{\begin{equation}}
\def\f{\end{equation}}
\def\E{\epsilon}

\newcommand{\p}{\partial}

\begin{document}

\title{Nonlinear-optical up and down frequency-converting backward-wave metasensors and metamirrors}

\author{Alexander K. Popov$^{1}$}%
\email{apopov@uwsp.edu}
\homepage{http://www4.uwsp.edu/physastr/apopov/}%
\author{
Igor. S. Nefedov$^2$}

\author{
Sergey A. Myslivets$^{3}$ }%
\author{
Mikhail I. Shalaev$^{4}$}
\author{
Vitaly V. Slabko$^{4}$}%

\affiliation{\vspace{1ex}$^1$\hbox{University of Wisconsin-Stevens Point, Stevens Point WI 54481, USA}\\
\hbox{and Birck Nanotechnology Center,
Purdue University,
West Lafayette, IN 47907, USA}\\
$^2$Aalto University, FIN-00076 Aalto, Finland,\\
\hbox{$^3$Institute of Physics, Siberian Branch of the
Russian Academy of Sciences,}\\  660036 Krasnoyarsk, Russian Federation,\\%
\hbox{$^4$Siberian Federal University,
660041 Krasnoyarsk, Russian Federation}}

\begin{abstract}
A concept of a family of unique backward-wave photonic devices, such as frequency up and down converting nonlinear-optical mirrors, sensors, modulators, filters and amplifiers  is proposed. Novel materials are considered, which support coexistence of ordinary and backward waves and thus enable enhanced nonlinear-optical  frequency-conversion processes. Particular properties of short-pulse regime are investigated.
\end{abstract}



  \maketitle

\section{INTRODUCTION}
\label{sec:intro}  
Optical electromagnetic radiation plays important role in pollution, chemical and image sensing. Hence, improvement of optical sensors is important for the indicated branches of sensing. This work is to further develop a concept of all-optically controlled, remotely actuated and interrogated sensor that can be employed for environmental probing in remote or hostile locations \cite{SPIEDSS11}. Extraordinary frequency-conversion processes in novel type of nonlinear-optical (NLO) materials are proposed and investigated that enable great enhancement and frequency conversion of signals, which carry important information, to the range of maximum sensitivity of detectors. Concurrently, the signals can be redirected towards the detectors. Particularly, extraordinary properties of coherent NLO energy exchange between ordinary and backward waves (BWs) in short pulse regime are investigated.

The possibilities of enhanced coherent energy exchange between electromagnetic waves and amplification of signals originate from  backwardness, the extraordinary property that electromagnetic waves acquire in negative index metamaterials (NIMs).  Unlike ordinary waves propagating in positive-index  materials, the energy flow, $\mathbf{S}$, and the wave-vector, $\mathbf{k}$,
become counter-directed in BWs, which determines their unique linear and NLO propagation properties. Usually, NIMs are nanostructured metal-insulator composites with a special design of their building blocks at the nanoscale.
Metal component  imposes strong absorption of optical radiation in NIMs, which presents a major obstacle towards their numerous prospective exciting applications. Extraordinary features of coherent NLO energy conversion processes in NIMs that stem from wave-mixing of ordinary and backward electromagnetic waves (BEMWs) and the possibilities to apply them for compensating the outlined losses have been predicted (for a review, see \cite{EPJD,SPIE} and references therein). Most remarkable feature is  distributed feedback behavior of output amplified and generated beams which allows for sharp resonance increase of the conversion efficiency for the appropriate intensity of fundamental beam. Essentially different properties of three-wave mixing (TWM) and second harmonic generation have been shown. Here, we propose two different classes of such materials: metamaterials (MM) with specially engineered spatial dispersion of their nanoscopic structural elements  and crystals that support optical phonons with negative group velocity. Both do not rely on nanoresonators which provide negative optical magnetism and constitute current mainstream in fabricating  NIMs. The appearance of BEMWs in metaslabs made of standing carbon nanotubes is shown. Uncommon phenomenon of generating of a contra-propagating wave at appreciably different frequency in the direction of reflection is investigated in the first class of MM.
The possibility to replace plasmonic NLO MMs, which are very challenging to fabricate, by the ordinary, readily available crystals is another proposed option. The possibility to mimic unparallel  NLO frequency-conversion propagation processes attributed to NIMs is shown for some of such crystals whereby optical phonons with negative group velocity and a proper phase-matching geometry are implemented. Here, optical phonons are employed instead of BEMWs.  The focus is on specific properties of  amplification and  nonlinear-optical energy exchange between the contra-propagating short pulses.

\section{Basic Idea}
\subsection{Huge Enhancement of Nonlinear Optical Energy Conversion and Signal Amplification through Three-wave Mixing of  Ordinary and Backward Electromagnetic Waves}
 An exotic electromagnetic property of NIMs stems from  the fact that energy flow and phase velocity of electromagnetic waves  become counter-directed inside the NIM slab.  Such phenomenon of {backwardness} of electromagnetic waves does not exist in naturally occurring materials.
The appearance of  BEMW can be explain as follows.
 The direction of the  wave-vector $\mathbf{k}$ with respect to the energy flow (Poynting vector)  depends on the signs of electrical permittivity $\epsilon$ and magnetic permeability $\mu $:
\begin{equation}
\mathbf{S}
=({c}/{4\pi})[\mathbf{E}\times\mathbf{H}] =({c^{2}\mathbf{k}}/{4\pi\omega\epsilon})H^{2}
s=({c^{2}\mathbf{k}}/{4\pi\omega\mu})E^{2}.  \label{s}
\end{equation}
If $\epsilon<0$ and $\mu<0$, refractive index becomes negative, $n= - \sqrt{\mu\epsilon}$, and vectors $\mathbf{S}$ and $\mathbf{k}$ become contradirected,
which is in striking contrast with the electrodynamics of ordinary, positive  index (PI) media (PIM).  Hence, magnetic response at optical frequencies, including magnetic nonlinear polarization,  opens new avenues in electromagnetics and  for its numerous revolutionary breakthrough applications. Such property does not exist in naturally occurring materials but becomes achievable in the plasmonic metamaterials.  Nonlinear-optical  propagation processes intrinsic to mixing of ordinary and backward electromagnetic waves have been shown to possess unprecedented properties \cite{1,2,SHG,3,4,6}.

\begin{figure}[h!]
\begin{center}
\resizebox{0.3\columnwidth}{!}{
\includegraphics{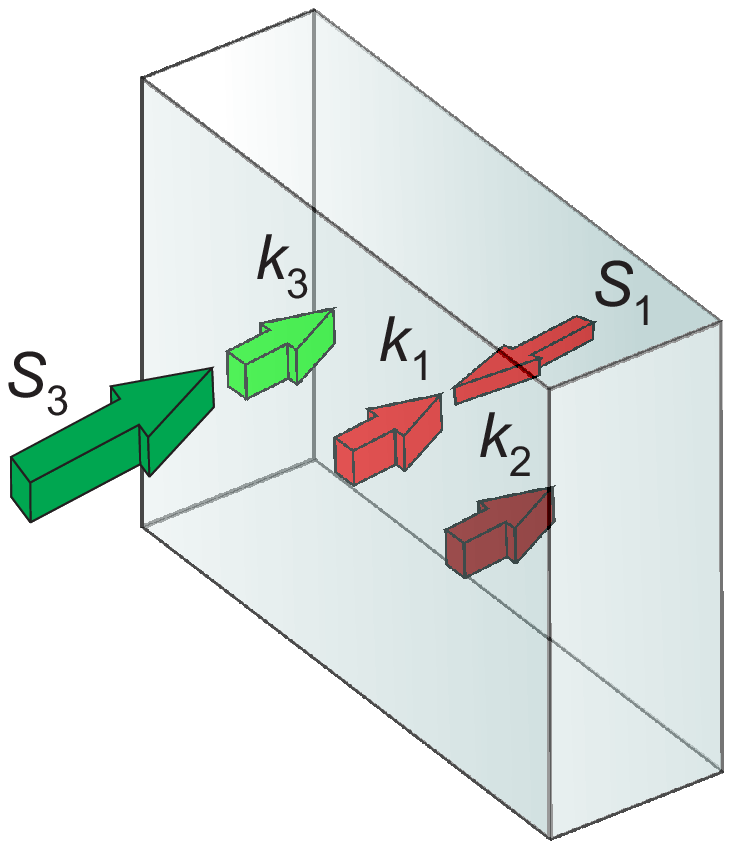}}
\resizebox{0.4\columnwidth}{!}{
\includegraphics{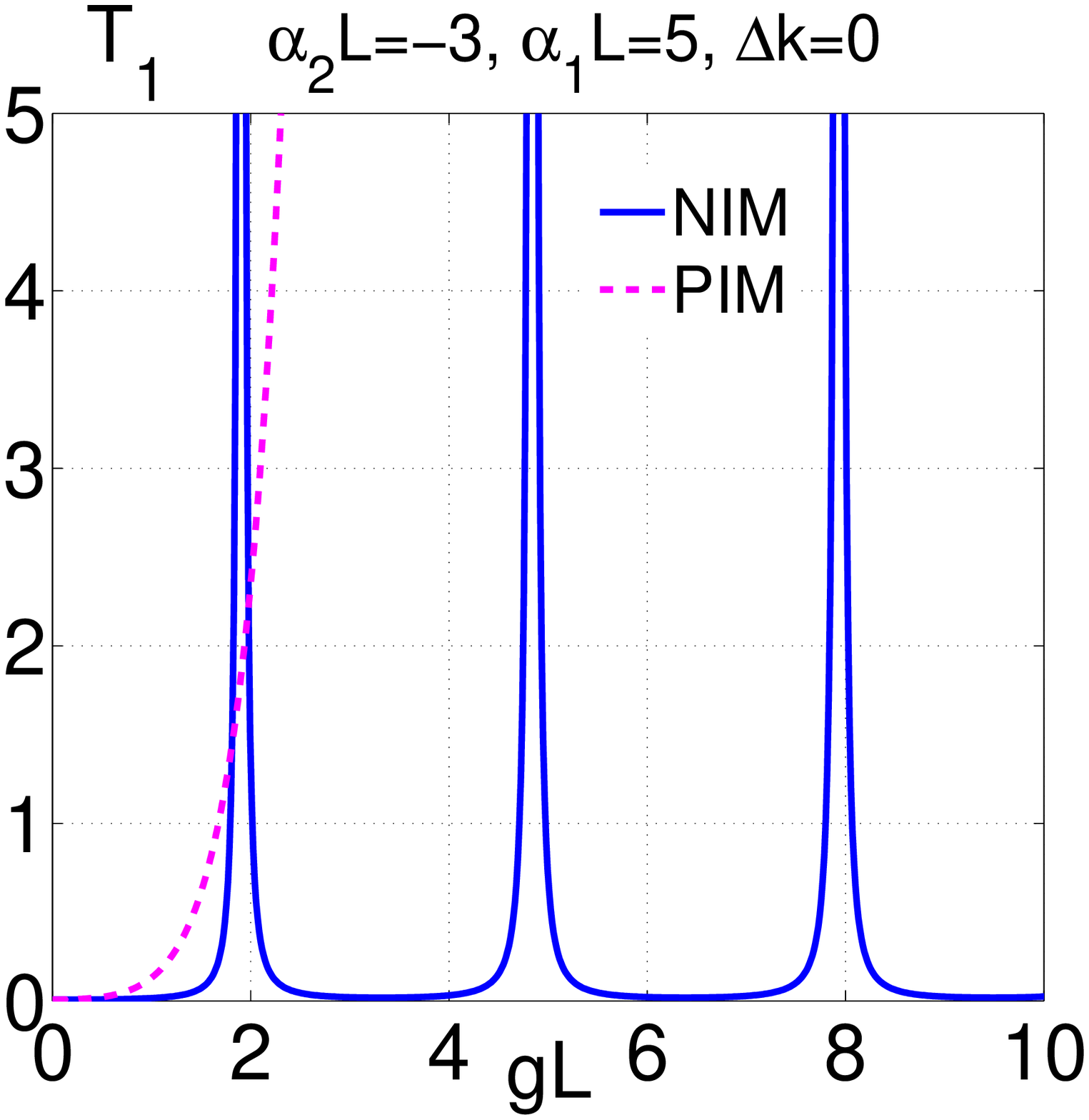}}\\
(a)\hspace{30mm}(b)\\
\resizebox{0.4\columnwidth}{!}{
\includegraphics{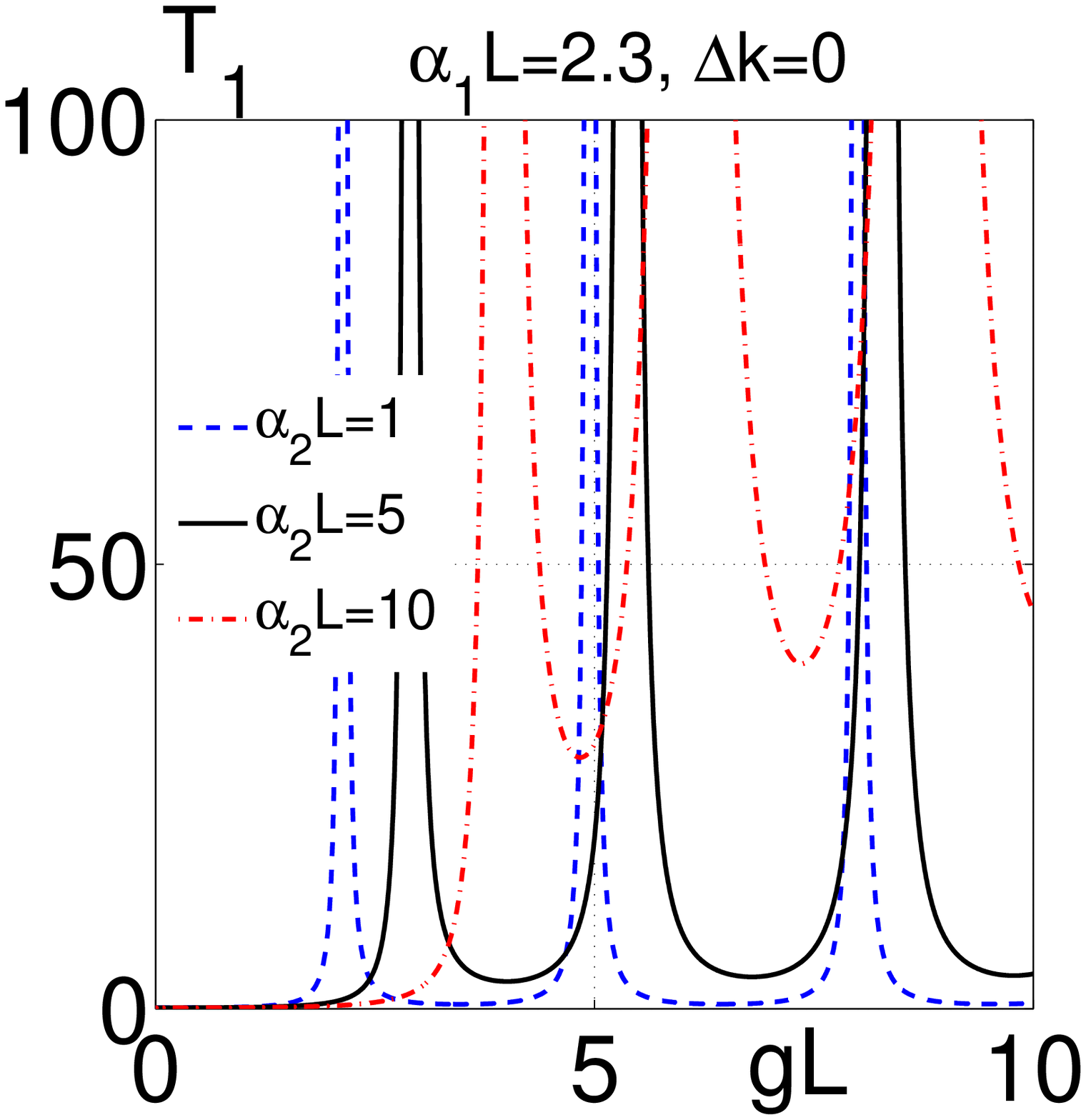}}
\resizebox{0.45\columnwidth}{!}{
\includegraphics{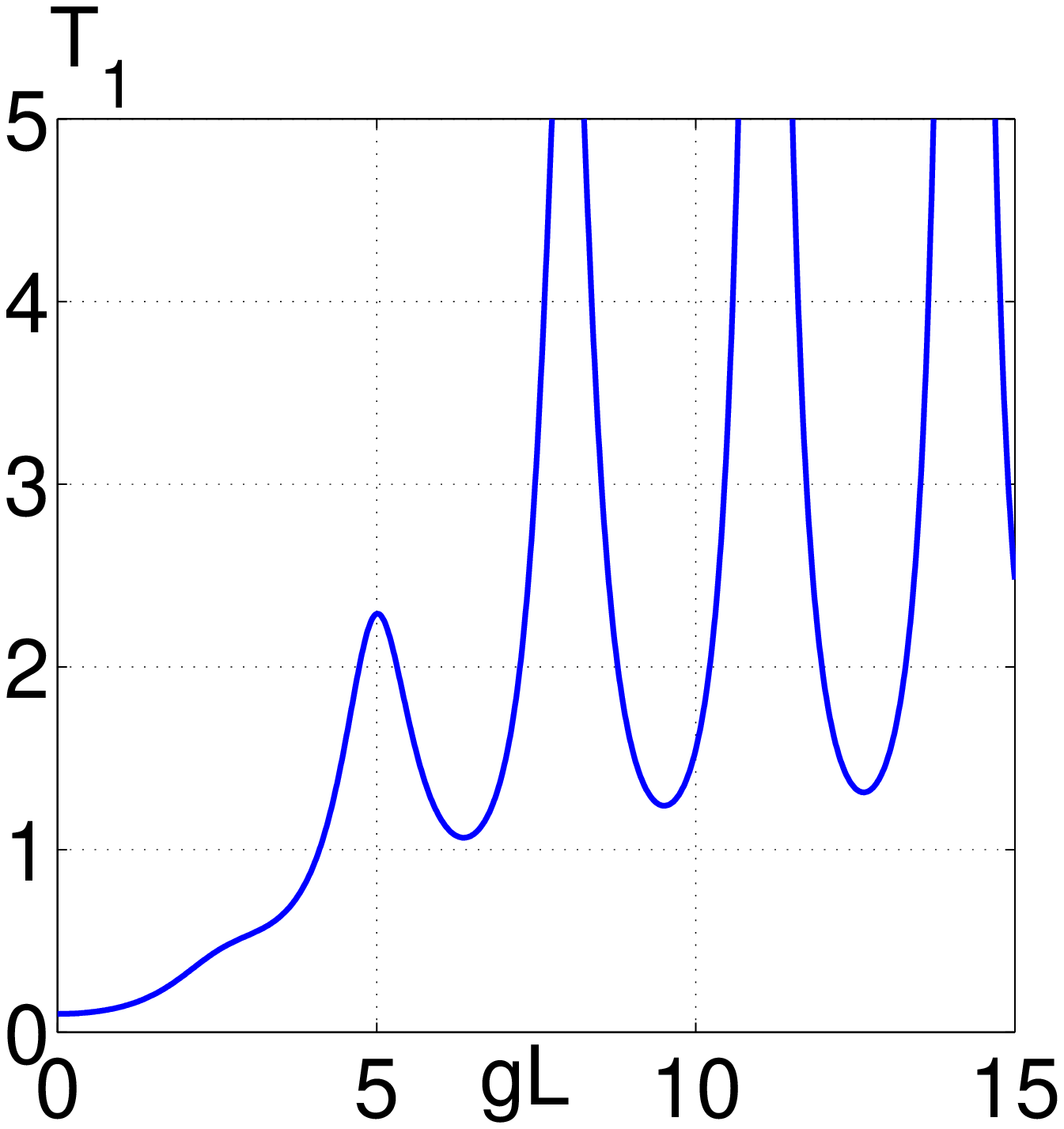}}
\\
(c)\hspace{30mm}(d)
\end{center}
\caption{\label{f21}(a) Coupling geometry. $\mathbf{S}_1$ -- negative index signal, $\mathbf{S}_3$ -- positive-index control field, $\mathbf{S}_2$ -- positive-index idler. (b) Solid: the dependence of the output signal $T_1(z=0)$ on the slab thickness and on intensity of the control field (on factor  $gL$). Here, the metamaterial is  absorptive at the frequency of the signal and assumed amplifying for the idler,  $\Delta k=0$. Dashed: the same dependence for the ordinary, PI media with the same other medium parameters.  (c) Transmission $T_1(z=0)$ of the signal at $\alpha_1L=2.3$ and different values of $\alpha_2L$. $\Delta kL=0$.
(d-f) Effect of phase-mismatch on the output signal. $\alpha_1L=2.3$, $\alpha_2L=3$, $\Delta kL=\pi$.  Diminished detrimental effect of phase mismatch on the transmission with the increase of intensity of the control field and/or the slab thickness is seen.}
\end{figure}

Unusual features of NLO coupling of the ordinary and backward EM waves that enable compensating losses for NI signal through optical parametric amplification (OPA) can be summarized as follows. Usually, plasmonic metamaterials possess NI properties only within a certain frequency band and ordinary, PI properties beyond such a band.  Figure~\ref{f21}(a) depicts  a slab of thickness $L$ with quadratic magnetic NLO response $\chi_m^{(2)}$. The medium with quadratic electric response would exhibit similar behavior. A NI (signal)  wave $H_1$ at $\omega_1$  enters the slab at its rear interface and a strong PI control field, $H_3$ at $\omega_3$, enters the slab at its front interface.  The control field at $\omega_3$  and weak signal at $\omega_1$ then generate a difference-frequency idler at $\omega_2=\omega_3-\omega_1$ which  falls in the PI frequency domain. The idler contributes back to the signal through the similar TWM process, $\omega_1=\omega_3-\omega_2$, and thus provides OPA of the signal. Such choice of the propagation directions ensures all wave vectors $\mathbf{k}_{1,2,3}$ to appear co-directed.

Crucial importance of the outlined geometrical resonances and striking difference of NLO propagation processes in negative index double domain NIM compared with their counterparts in ordinary materials are illustrated in Fig.~\ref{f21}  adopted from our paper \cite{SPI}. Besides the factor $g$, the local NLO energy conversion rate for the signal is proportional to the amplitude of the idler (and vice versa)  and depends on the  phase mismatch $\Delta k$. Hence, the fact that the waves decay in opposite directions causes a specific, strong dependence of the entire propagation process and, consequently, of the transmission properties of the slab on the {ratio} of the signal and the idler decay rates.
Such extraordinary resonance behavior,  which occurs due to the backwardness of the light waves in NIMs, is explicitly seen when compared with similar distribution in ordinary, PI materials depicted in Fig.~\ref{f21}(b).
Basically, such induced transparency resonances are narrow, like those depicted in Fig.~\ref{f21}(b) and by the plot in Fig.~\ref{f21}(c) corresponding to $\alpha_2L=1$. This indicates that the sample remains opaque anywhere beyond the resonance values of the control field and of the sample thickness. Any sharp frequency dependence of nonlinear susceptibility or absorption indices  translates into frequency \emph{narrow-band} filtering.
The slab  becomes \emph{transparent} within the broad range of the slab thickness and the control field intensity if the transmission in all of the minimums is about or more than~1. Figures~\ref{f21}(c,d)  show the feasibility of achieving {robust transparency and amplification in a NIM slab at the signal frequency through a wide range of the control field intensities and slab thicknesses} by the appropriate adjustment of the absorption indices $\alpha_{2}\geq\alpha_{1}$.
Oscillation amplitudes grow sharply near the resonances, which indicates the possibility of  \emph{cavity-less
self-oscillations}. The distribution of the signal and the idler inside the slab would also dramatically change.  These simulations prove that, unless optimized, the signal maximum inside the slab may appear much greater than its output value at $z=0$.
Giant enhancement in the resonances indicates that strong absorption  of the left-handed, negative-phase wave and of the idler can be turned into transparency, amplification and even into cavity-free self-oscillation.   Self-oscillations would provide for the generation of \emph{entangled counter-propagating}
left-handed, $\hbar\omega_1$, and right-handed, $\hbar\omega_2$,
photons without a cavity.
The outlined features can be employed for design of ultracompact optical sensors, selective filters, amplifiers and oscillators generating  beams of {counterpropagating entangled} photons.

\subsection{Three Alternative Coupling Schemes -- Three Sensing Options }

\begin{figure}[h!]
\begin{center}
\includegraphics[width=.25\columnwidth]{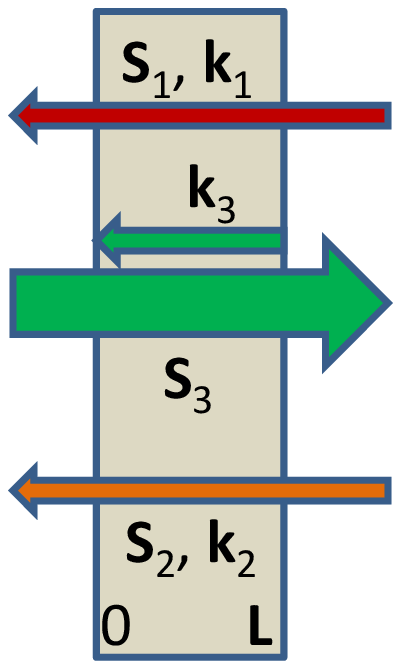}
\includegraphics[width=.25\columnwidth]{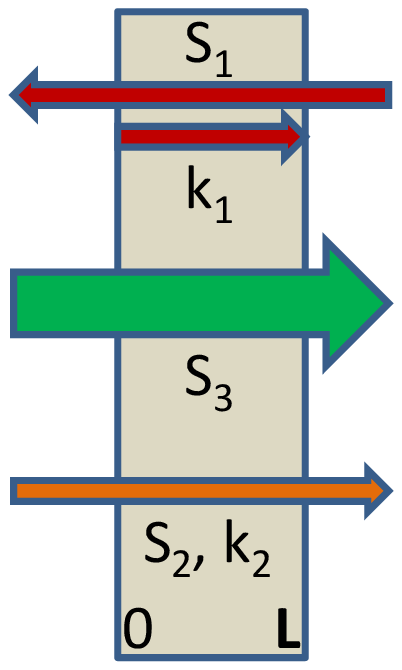}
\includegraphics[width=.25\columnwidth]{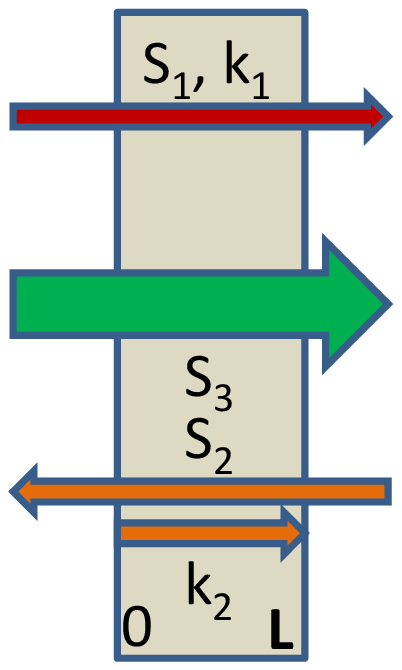}\\
(a)\hspace{20mm}(b)\hspace{20mm}(c)
\end{center}
\caption{\label{fig1} Three different options of the proposed NLO sensors. $\mathbf{S}_{1,2}$ and $\mathbf{k}_{1,2}$  are energy fluxes and wavevectors  for the ordinary, positive index, signal and generated idler; $\mathbf{S}_3$ and $\mathbf{k}_3$  -- for the negative index control field.
(a) Frequency-converting metaamplifier.  (b) and (c) – frequency-converting and amplifying metamirrors. (b) The NLO sensor amplifies the signal ${S}_1$ travailing against the control beam [$n(\omega_1) < 0$],   frequency up-converts   ($S_2$)  and sends it  along  the  control beam $S_1$. (c) The NLO sensor  amplifies the signal ${S}_{1}$ traveling along the control field, up-converts its frequency of and sends it back in the direction against the control beam [$n(\omega_2)<0$].}
\end{figure}

 Figure~\ref{fig1} depict three possible options for the phase matched  NLO coupling of the ordinary and backward waves.
Consider the example depicted in panel~(c). Assume that the  wave at $\omega_{1}$ with the wave-vector $\mathbf{k}_1$ directed along the $z$-axis is a PI ($n_{1}>0$) signal. Usually it experiences strong absorption caused by metal inclusions. The medium is supposed to possess a quadratic nonlinearity $\chi^{(2)}$ and is illuminated by the strong higher frequency control field at $\omega_{3}$, which also falls into the PI domain. Due to the  three-wave mixing (TWM) interaction, the control and the signal fields generate a difference-frequency idler at $\omega_{2}=\omega_{3}-\omega_{1}$, which  is  a NI wave ($n_{2}<0$). The idler, in cooperation with the control field, contributes back into the wave at $\omega_{1}$ through the same type of TWM interaction and thus enables optical parametric amplification (OPA) at $\omega_{1}$  and enhanced reflected beam at $\omega_{2}$  by converting the energy of the control fields into the signal and the idler.  In order to ensure effective energy conversion,  the traveling wave of nonlinear polarization of the medium and the coupled electromagnetic wave at the same frequency must be phase matched, i.e., must meet the requirement of $\Delta \mathbf{k} = \mathbf{k}_3-\mathbf{k}_2-\mathbf{k}_1=0$. Hence, all phase velocities (wave vectors) must be co-directed. Since $n(\omega_2)<0$, the idler  is a BW, i.e., its energy flow $\mathbf{S}_{2}=(c/4\pi)[\mathbf{E_2}\times \mathbf{H_2}]$ appears directed against the $z$-axis. This allows to conveniently remotely interrogate the NLO  chip and to actuate amplification and frequency up-conversion of  the signal. Transmitted and upconverted waves are sent  in the opposite directions towards the remote detectors. Such a signal can be, e.g., infrared thermal radiation emitted by the object of interest, or signal that carries important spectral information about the chemical composition of the environment. The  schemes depicted in Fig.~\ref{fig1}(a),(b) offer two other options with different operational properties for nonlinear-optical sensing. The research challenge is that such unprecedented NLO coupling schemes lead to changes in the set of coupled nonlinear propagation equations and boundary conditions as compared with the standard ones known from the literature. This, in turn,  results in dramatic changes in their solutions and in multiparameter dependencies of the operational properties of the proposed sensor.

\section{Coherent Nonlinear Optical Coupling of Ordinary and  Backward Electromagnetic Waves in Metamaterial made of Carbon Nanotubes}
Exciting unparallel avenues for nonlinear electromagnetics can be open  by the  metamaterials  where formation of BWs becomes possible due to  specific spatial dispersion of their structural elements.
Basic idea is as follows.  As outlined above, according to currently commonly adopted concept, negative refractive index and associated backwardness of optical waves require negative permeability and therefore magnetism at optical frequencies.
 However, a different approach is possible \cite{Agr,AgGa}.  In a loss-free isotropic medium, energy flux $\mathbf{{S}}$ is  directed along the group velocity ${\mathbf{v}_g}$:
${\mathbf{S}}={\mathbf{v}_g}U$, ${\mathbf{v}_g}=\rm grad_{\mathbf{k}}\omega(\mathbf{k})$. 
Here, $U$ is energy density attributed to EMW. It is seen that the group velocity   may become directed \emph{against} the wavevector depending on sign of dispersion $\partial\omega/\partial k$. Basically, negative dispersion can appear in fully dielectric materials with particular dispersion of its structural elements.
This opens an entirely novel research and application avenue. In the next subsection, we show a possibility to engineer such a dispersive  medium which supports  coexistence of both ordinary and BEMWs which can be phase matched. An example of enhanced coherent energy exchange between ordinary fundamental EMW and its second harmonic is considered in a lossless NLO slab for the sake of simplicity.

\subsection{Carbon ``Nanoforest'' and Phase Matching of Ordinary and Backward EM Waves.}

Appearance of BEM modes in nanoarrays and layered structures has been predicted recently in Ref. \cite{Nef,Bel,IgorPRB}. Obviously, many other options should have existed.
Below, we propose such an option that seems promising in the context of \emph{nonlinear} propagation and coherent energy conversion processes. Namely, the possibility of conversion of ordinary EMW to the contrapropagating BEMW at its doubled frequency. Hence, such metaslab can be viewed as a frequency-doubling NLO metamirror.

\begin{figure}[h]
\begin{center}
\includegraphics[width=.5\columnwidth]{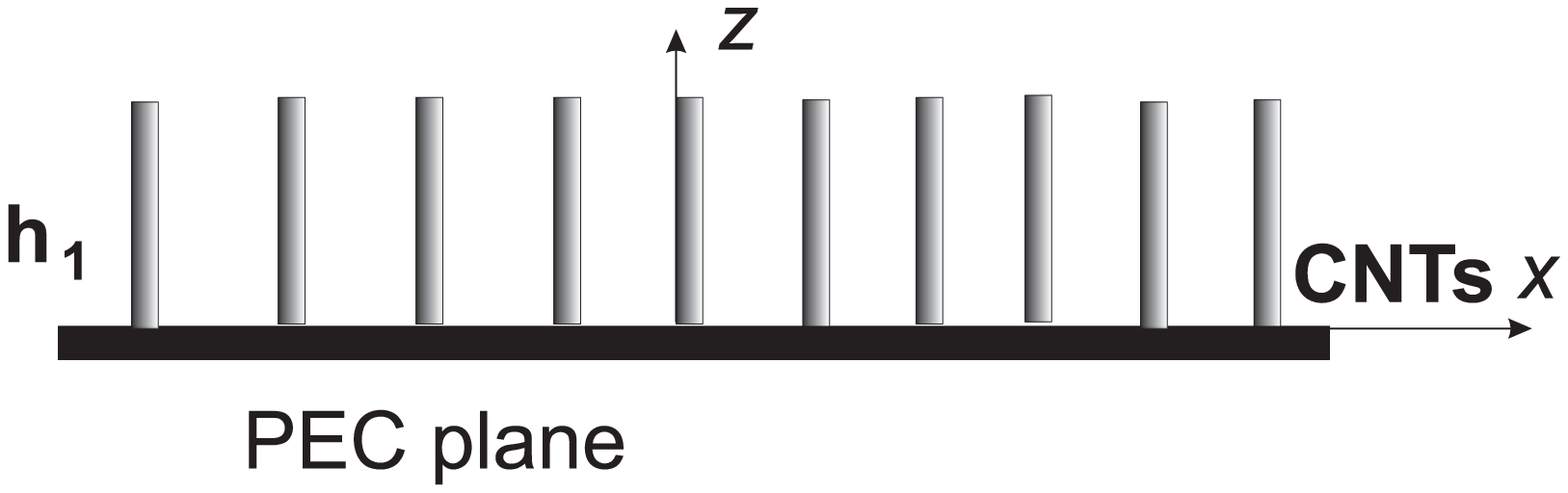}\\
(a)\\
\includegraphics[width=.5\columnwidth]{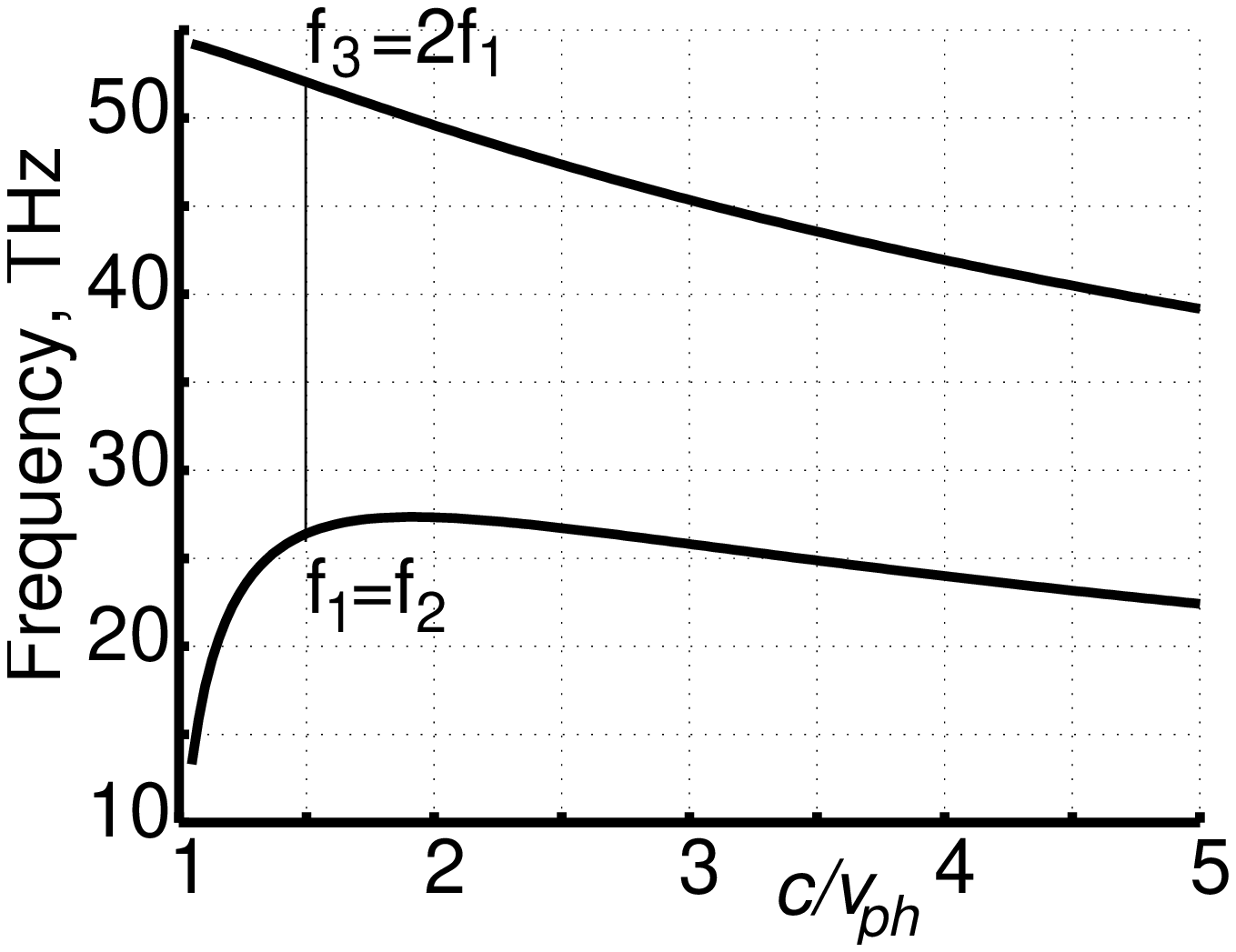}\\
(b)\\
\includegraphics[width=.5\columnwidth]{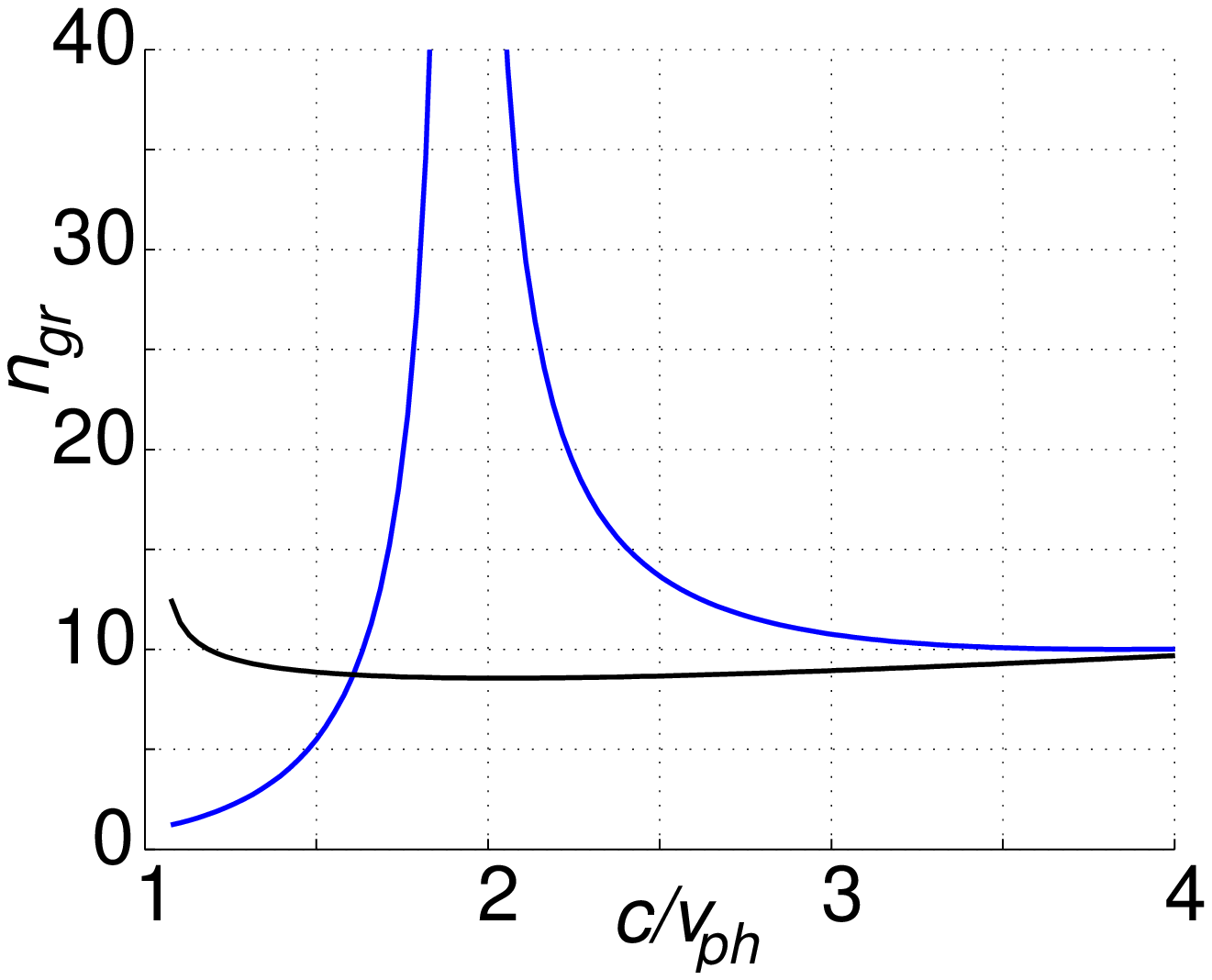}\\
(c)
\caption{\label{CNT}  (a) Geometry of free-standing CNTs. (b) Dispersion -- the frequency vs slow-wave factor  for the slab of CNTs with open ends. (c) Group delay factor vs the phase velocity slow-wave factor for the same modes as in panel (a). Black (flat) curve corresponds to the high-frequency mode, blue curve to the low-frequency mode. The tip of blue curve is cut. Its maximum corresponds to the stop-light regime. }
\end{center}
\end{figure}
Figure \ref{CNT}(a) depicts a periodic array of carbon nanotubes (CNT) vertically standing on the surface of a perfect electric conductor (PEC) with the CNT ends open to air, which can be seen as perfect magnetic conductor (PMC).
As  shown in Ref.  \cite{IgorPRB},  EM waves travelling through such CNT ``nanoforest,'' along x or y directions, posses a  hyperbolic dispersion and relatively low losses in the THz and mid-IR ranges. One of the most important consequences from the hyperbolic-type dispersion law is the possibility for  propagation of both forward \emph{and backward} EM waves.
Consider EMW propagating along the x-axis.
We also introduce  group $n_{gr}=c/v_g$ and phase $n_{ph}=c/v_{ph}$ slow-wave factors. The latter one is refractive index.
For the given case of  surface waves propagating in the slab of CNTs with open ends, whose fields attenuate in air, the dispersion
is given by the equation:
\e\label{dis}
\tan{(k_zh)}={\sqrt{k_x^2-k^2}}/{k_z}.
\f
Such a dependence can be understood  from considering a planar waveguide formed by perfect PEC  and PMC planes and tampered with a CNT array. The array axis is orthogonal to the walls of the waveguide.
Then, the propagation constant along the waveguide is
$k_{\perp}=\sqrt{\E_{zz}\left[k^2-(m\pi/2h)^2\right]}$, 
where $m$ is a positive integer,  $h$ is the height of the waveguide (CNT) and $k$ is the wavenumber in free space.
If $\E_{zz}<0$, BW propagation is allowed
when $k<m\pi/2h$ and forbidden for $k>m\pi/2h$.
The relation between the  wavevector component $k_x$ and wavenumber  $k$ is:
$k_x^2=[{(k^2-k_p^2)(k^2-k_z^2)}]/[{k^2}]$, 
where $k_z=m\pi/(2h)$, $m$ is the integer determining a number of field variations along CNT, $k_p$ is plasma wavevector.
One can  show that  ${\rm d}k_{\perp}^2/{\rm d}k^2<0$, if $k_z/k>1$ and $k_p/k>1$.

Numerical analysis of  Eq.~(\ref{dis}) is depicted in Fig.~\ref{CNT}(b) for the case of  CNT radius $r=0.82$~nm,  the lattice period $d=15$~nm and EM modes with $m=1$ and $m=3$. Slow-wave factor is proportional to the wave vector: $c/v_{ph}=k(\omega)/k_0$, where $k_0$ is the wave vector in the vacuum. The appearance of positive dispersion for small slow-wave factors is caused by interaction of BW in the CNT slab with the plane wave in air. Indeed, coexistence of the positive (ascending dependence) and negative (descending dependence) dispersion for different frequencies proves that such a metamaterial  supports both ordinary and \emph{backward} EMW. It also proves that resonant plasmonic structures, like split-ring resonators, exhibiting negative $\epsilon$ and $\mu$ are \emph{not} the necessary requirement for the realization of BW regime in mid-IR range. The possibility of considerable increased bandwidth of BEMW  compared to most plasmonic MM  made of nanoscopic resonators is seen that gives  the ground to consider CNT arrays as a promising \emph{perfect backward-wave metamaterial}. The slow-wave factor for both modes is shown in Fig.~\ref{CNT}(c). The magnitude of $n_{gr}$ goes to infinity at $n_{ph}\approx 1.85$, which indicates the stop-light regime for the low-frequency mode. Particularly, Fig.~\ref{CNT}(b) shows the possibility of \emph{phase matching} of ordinary fundamental and backward second harmonic EMWs.
\subsection{Coherent Energy Exchange Between Short Contrapropagating Pulses in the Carbon Nanoforest}
Here, we demonstrate unusual dependence of pulse shape and overall efficiency of SHG on the ratio of input fundamental pulse length and metamaterial slab thickness. Consider a double domain positive/negative index slab of thickness $L$ that supports ordinary EMW at fundamental frequency (FH) and BEMW at SH frequency. To ensure phase matching, wavevectors of the FH and SH must be co-directed and, hence, their energy fluxes - counter-directed. The slab operates as a \emph{frequency up-converting nonlinear-optical mirror} with controllable reflectivity. The equations for amplitudes of FH, $a_1$, and SH, $a_2$, can be written as
\begin{eqnarray}\label{eq1}
 \frac1{v_1}\frac{\p a_1}{\p t}+ \frac{\p a_1}{\p z}= -i2ga_1^*a_2\exp{(i\Delta kz)}-\frac{\alpha_1}2a_1,\\ -\frac1{v_2}\frac{\p a_2}{\p t}+ \frac{\p a_2}{\p z}= iga_1^2\exp{(-i\Delta kz)}+\frac{\alpha_2}2a_2.
\end{eqnarray}
Here, $|a_{1,2}|^2$ are slowly varying amplitudes proportional to the instant photon numbers in the energy fluxes, $\alpha_{1,2}$ are absorption indices, $\Delta k=k_2-2k_1$ is phase mismatch, and $v_i$ are the group velocities for the corresponding pulses.
\begin{figure}[ht!]
\begin{center}
\includegraphics[width=.49\columnwidth]{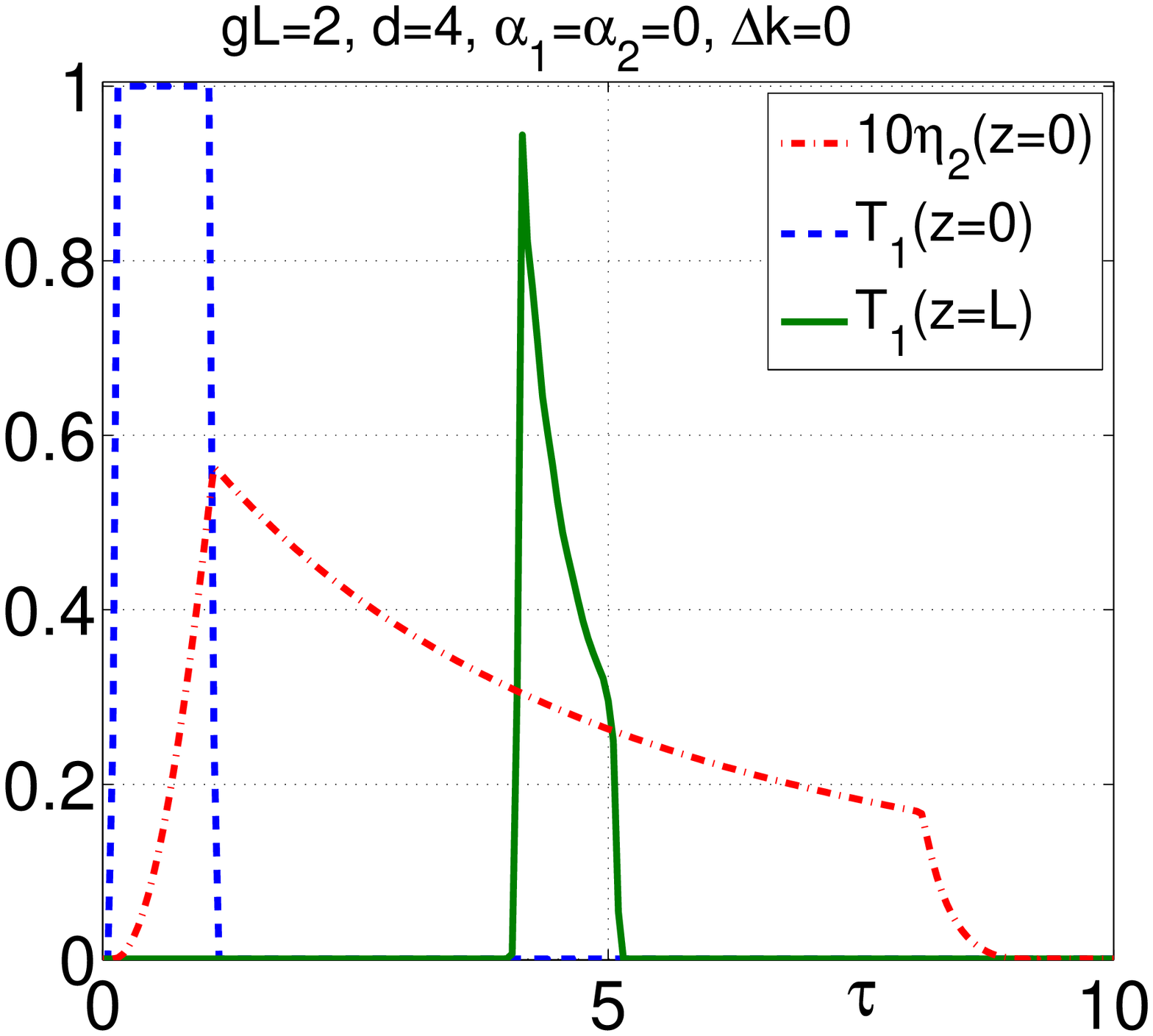}
\includegraphics[width=.49\columnwidth]{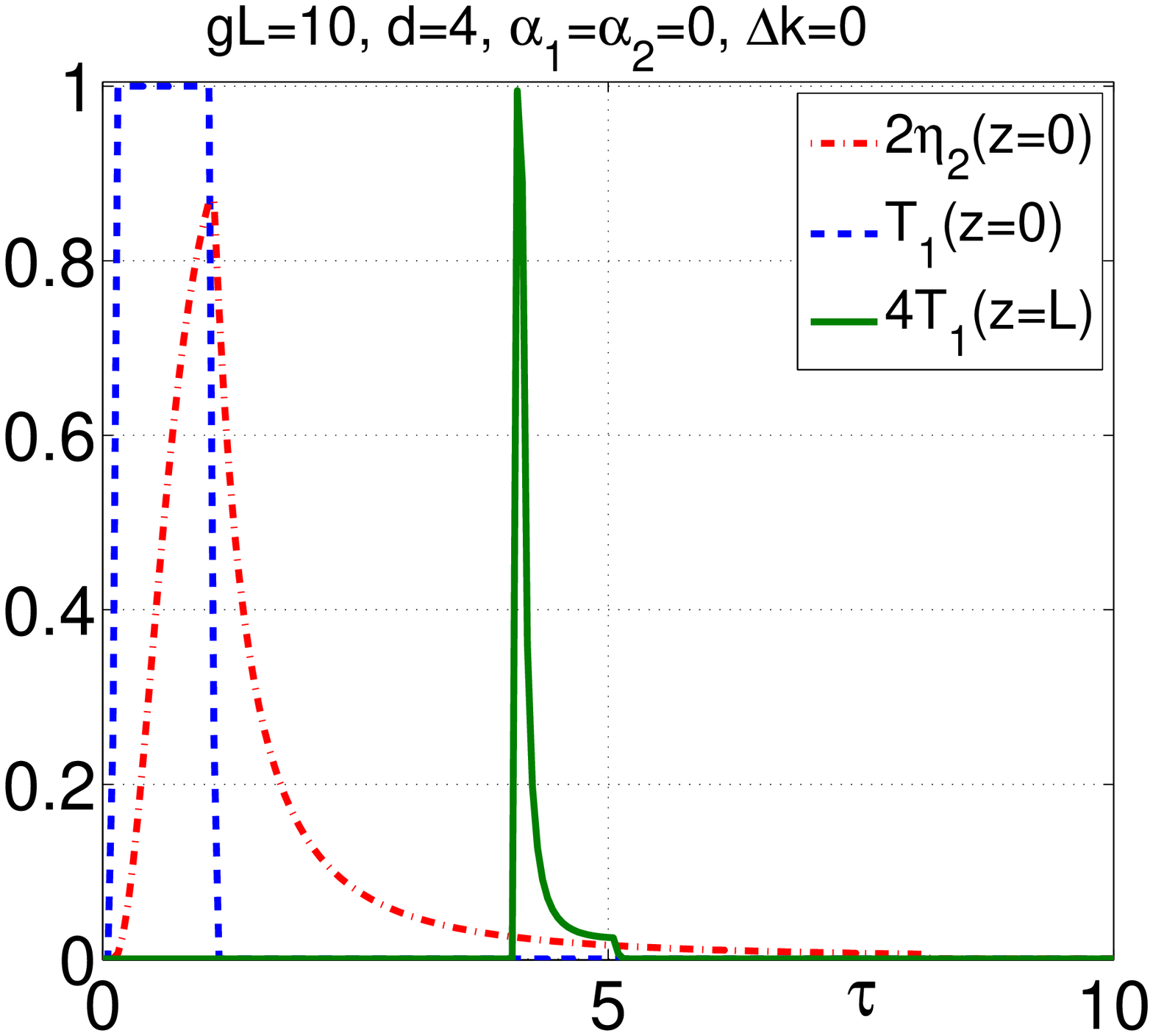}\\
(a)\hspace{35mm} (b)\\
\includegraphics[width=.49\columnwidth]{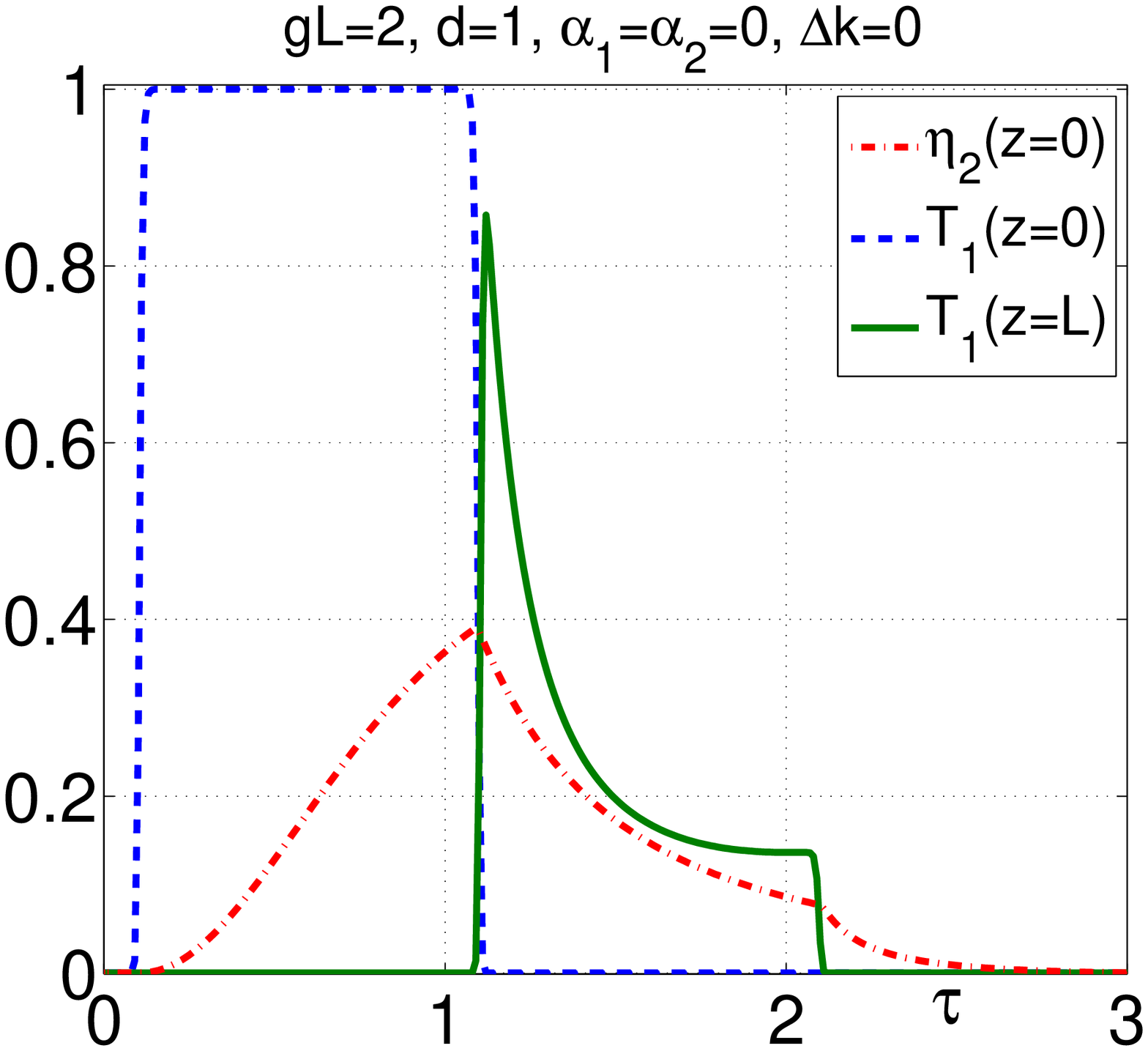}
\includegraphics[width=.49\columnwidth]{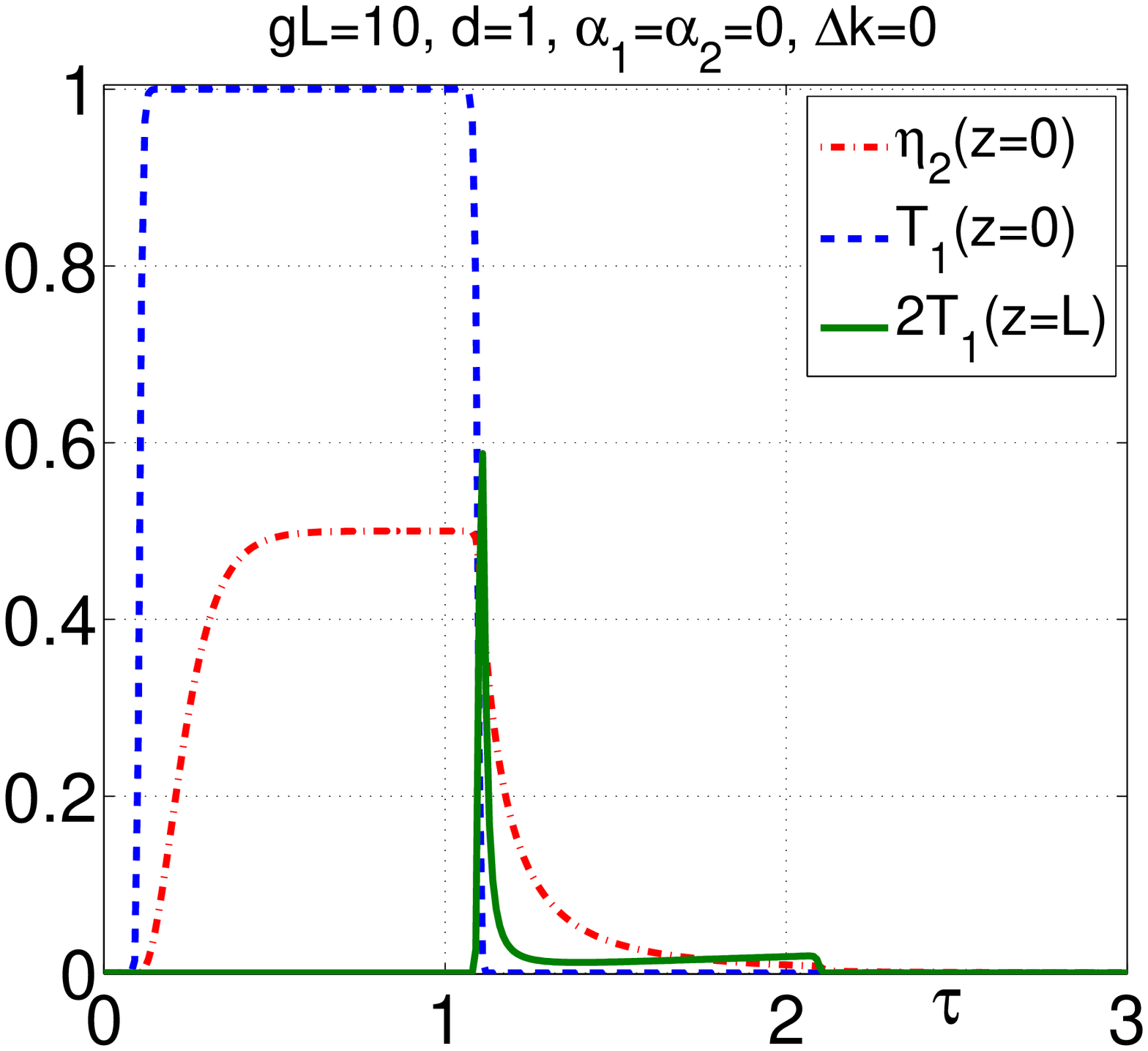}\\
(c)\hspace{35mm} (d)
\end{center}
\caption{\label{fi1} \small Input, $T_1(z=0)$,  and output, $T_1(z=L)$,  pulses of fundamental and negative-index SH radiation,  $\eta_2(z=0)$. (a) and (b) -- short pulse: d=4. (c) and (d): longer pulse: d=1.  (b) and (d) -- input power is 25 times larger than in (a) and (c).
(a) Input pulse area (energy) $S_{10}=0.9750$; output pulse areas (energy)  $S_{1L}= 0.5031$, $S_{20}= 0.2392$. (b)  Input pulse energy $S_{10}=0.9750$; output pulse energy $S_{1L}= 0.0396$, $S_{20}= 0.4742$. (c) Input pulse energy $S_{10}= 0.9900$; output pulse energy $S_{1L}=  0.2516$, $S_{20}= 0.3692$. (d)  Input pulse energy $S_{10}= 0.9900$; output pulse energy $S_{1L}=0.0161$, $S_{20}= 0.4870$.
}
\end{figure}
Note opposite signs in the equations and the requirement of the boundary conditions to be set at the opposite edges of the slab for FH and SH. These lead to cardinal changes in the solutions to the equations as compared to those in the ordinary nonlinear optical material.  We have chosen the input pulse shape as being close to a rectangular form
\begin{equation}
F(\tau)=0.5\left(\tanh\frac{\tau_0+1-\tau}{\delta\tau}-\tanh\frac{\tau_0-\tau}{\delta\tau}\right),
\end{equation}
where $\delta\tau$ is the duration of the front and tail, and $\tau_0$ is the shift of the front relative to $t=0$. All quantities are reduced by the pulse duration $\Delta \tau$. The magnitudes $\delta\tau=0.01$ and $\tau_0=0.5$ have been selected for numerical simulations.

Unusual properties of BWSHG in NIMs in the pulsed regime stem from the fact that it occurs only inside the traveling pulse of FH. SHG begins on its leading edge, grows towards the back edge, and then exits the pulse with no further changes. Since the FH pulse propagates across the slab, the duration of the SH pulse is longer than the fundamental one. Depletion of the FH radiation along the pulse length and the conversion efficiency depend on its initial maximum intensity and phase matching. Ultimately, the overall properties of SHG, such as the pulse length, and the photon conversion efficiency, appear dependent on the ratio of the FH pulse and slab lengths. Such an extraordinary behavior is illustrated in Figs.~\ref{fi1}(a)-(d).
Here,  $d$  is the slub thickness reduced by the input pulse length, $d=L/v_1\Delta\tau$,  $g$ is proportional to product of nonlinear susceptibility $\chi^{(2)}$ and the input amplitude of FH.
A rectangular shape of the  input FH pulse $T_1=|a_1(z)|^2/|a_{10}|^2$ is depicted at $z=0$ when its leading front enters the medium.  The results of numerical simulations  for the output FH pulse,  when its tail passes the slab's edge at $z=L$, as well as for the shape and conversion efficiency of the output  SH pulse,  $\eta_2=|a_2(z)|^2/|a_{10}|^2$,
when its tail passes the slab's edge at $z=0$, are shown. For clarity, here, the medium is assumed loss-free,  group velocities of the fundamental and SH pulses assumed equal, $\Delta k=0$.
Panels (a) and (b) correspond to the fundamental pulse four time shorter than the slab thickness. Increase of the conversion efficiency  with increase of the intensity of the input pulse is seen. It is followed by shortening of the SH pulse.
The outlined properties satisfy to the conservation law: the number of annihilated pair of photons of FH radiation ($S_{10}-S_{1L})/2$ is equal to the number of output SH photons $S_{20}$.
Panels (c) and (d) display corresponding changes for a longer input pulse with the length equal to the slab thickness. Here, larger conversion efficiency can be achieved at a lower input intensity compared with the preceding case because of the longer conversion length. The changes in the SH pulse length and conversion efficiency with increase of input intensity appear less significant.

\section{ Fully Dielectric  Backward-Wave NLO Material: Enhancing Coherent Energy Transfer Between Electromagnetic Waves in Ordinary Crystals by Coupling with Optical Phonons with Negative Phase Velocity\label{ph}
}

As above described, NLO with backward \emph{electromagnetic} waves enables a great enhancement of energy-conversion rate at the otherwise  equal nonlinearities and intensities of input waves. Herein, we propose \emph{fundamentally different} scheme of TWM of ordinary and backward waves (BW). It builds on the stimulated Raman scattering (SRS) where two ordinary EM waves excite backward elastic vibrational wave in a crystal, which results in TWM. The possibility of such BWs was predicted by L. I. Mandelstam in 1945 \cite{Ma}, who also had pointed out that negative refraction is a \emph{general property} of the BWs. The idea underlying the proposed concept and its basic justification is described below. The goal is to show the possibility to replace the NI plasmonic composites, which are challenging to fabricate,  with readily available ordinary crystals, some of which have been already extensively studied, and thus to mimic the unparallel properties of coherent NLO energy exchange between the ordinary and BW.
The basic idea is as follows. In \cite{ph}, stimulated Raman scattering (SRS) on optical phonon  was investigated in continuous wave (CW) regime. The possibility of distributed-feedback type resonance enhancement of amplification has been shown that stem from negative dispersion  of elastic waves. The effect appeared similar to that in three-wave mixing (TWM) of ordinary and BEMWs in NIMs, provided that special requirements are met.  Yet, the required intensity of the fundamental field was found to be close to the optical breakdown threshold due to the high phonon damping rate. Here, we show that the indicated fundamental formidable obstacle  can be removed by making use of short pulses, which opens the possibility to mimic TWM processes attributed to NIMs in readily available crystals. Besides that, we show that such mixing exhibits unparallel properties that allow for tailoring the shapes of the  generated and transmitted pulses and for huge enhancement of the frequency conversion efficiency.

\begin{figure}[h]
\begin{center}
\includegraphics[width=.7\columnwidth]{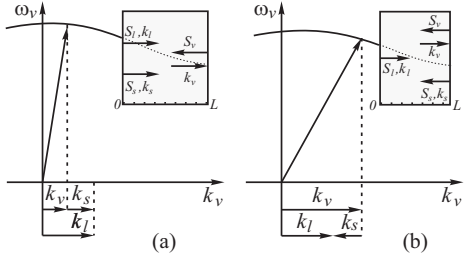}
\caption{\label{phf1}  Negative dispersion of optical phonons and two phase matching options for short- and long-wave vibrations: (a) -- co-propagating, (b) -- contra-propagating  fundamental (control)  and Stokes (signal) waves. Insets: relative directions of the energy flows and the wave-vectors. } \end{center}
\end{figure}
\begin{figure}[t]
\begin{center}
\includegraphics[width=.42\columnwidth]{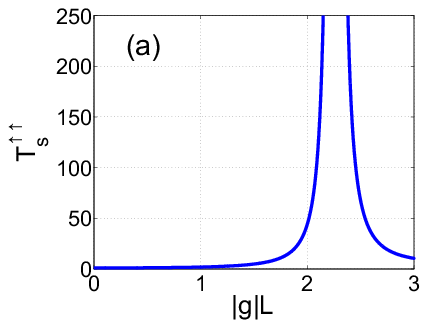}
\includegraphics[width=.37\columnwidth]{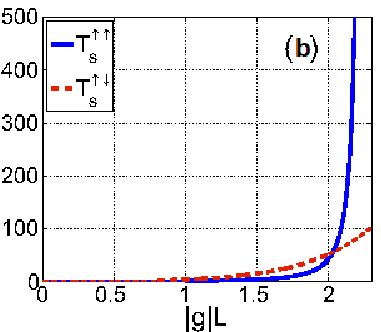}
\caption{\label{phf2}
 (a) Transmission of the Stokes wave $T_s^{\upuparrows}(z=L)$ vs intensity of the fundamental control field in the vicinity of  first ``geometrical'' resonance [co-propagating $E_l$ and $E_s$ geometry, Fig.~\ref{phf1}(a)]. $g=\sqrt {g_v^* g_s }$. Such extraordinary resonance appears because of backwardness of the coupled vibration wave and opposite direction of propagation for the Stokes and phonon waves.
(b)~Comparison of the output intensities of the Stokes wave vs intensity of the control field for co- (the blue, solid line) and contra-propagating (the red, dashed line)  fundamental (control) and signal (Stokes) waves [coupling geometries depicted in Fig.~\ref{phf1}(a) and (b) respectively].}
\end{center}
\end{figure}

Typical dispersion curve $\omega(k)$ for optical phonons, which exist in crystals  containing more than one atom per unit cell, is depicted in Fig.~\ref{phf1}. The dispersion is negative  in the range from zero to the boundary of the first Brillouin's zone. Hence, the group velocity of such phonons $\mathbf{v}_{v}$ is antiparallel with respect to its wave-vector $\mathbf{k}_{v}$. The dispersion $\omega _v(k_v)$ can be approximated as
$\omega _v =\sqrt {\omega _0^2 -\beta k_v^2 }$.
Then,  in the vicinity of  $k_v=0 $, velocity $v_v^{gr}$  is given by
$v_v^{gr} =-\beta {k_v }/{\omega _v }=-{\beta }/{v_v^{ph}}$,
where $v_v^{ph} $ is the projection of the phase velocity  of the vibrational wave on the z-axis and $\beta $ is the  dispersion parameter for the given crystal.
eq1Optical elastic vibrations can be excited by the light waves through the Raman scattering. The latter gives the ground to consider such a crystal as the analog of the medium with negative refractive index at the phonon frequency and to employ the processes of parametric interaction of three waves, two of which are ordinary EM  waves and the third one is the backward wave of elastic vibrations.
The  coupled waves are described by the equations
\begin{eqnarray} \label{eq1a}
 E_{l,s} &=&({1}/{2})\mathcal{E}_{l,s} (z,t)e^{ik_{l,s} z-i\omega _{l,s}
t}+c.c. ,\\
 Q_v &=&({1}/{2})Q(z,t)e^{ik_v z-i\omega _v t}+c.c.
\end{eqnarray}
Here, $\mathcal{E}_{l,s} $,  $Q$, $\omega _{l,s,v} $ and $k_{l,s,v} $ are the amplitudes, frequencies and wave-vectors of the  fundamental, Stokes and vibrational waves;  $Q_v (z,t)=\sqrt \rho x(z,t)$; $x$ is displacement of the vibrating particles,  $\rho $ is the medium density and the requirements
$ \omega _l =\omega _s +\omega _v \left( {k_v } \right) $,\quad
$ \vec {k}_l =\vec {k}_s +\vec {k}_v$
are supposed met. Partial differential equations for the slowly varying amplitudes in the approximation of the first order of $Q$ in the polarization expansion are \cite{Bl}:
\begin{eqnarray}
 \frac{\partial \mathcal{E} _l }{\partial z}+\frac{1}{v_l}\frac{\partial \mathcal{E} _l }{\partial t}&=&i\frac{\pi \omega _l^2 }{k_l
c^2}N\frac{\partial \alpha }{\partial Q}\mathcal{E} _s Q \label{l}\\
 \frac{\partial \mathcal{E} _s }{\partial z}+\frac{1}{v_s
}\frac{\partial \mathcal{E} _s }{\partial t}&=&i\frac{\pi \omega _s^2 }{k_s
c^2}N\frac{\partial \alpha }{\partial Q}\mathcal{E} _l Q^\ast \label{s1}\\
 \frac{\partial Q}{\partial z}-\frac{1}{v_v }\frac{\partial Q}{\partial
t}-\frac{Q}{v_v\tau_v}&=&-i\frac{1}{4\omega _v v_v }N\frac{\partial
\alpha }{\partial Q}\mathcal{E} _l \mathcal{E} _s^\ast.\label{q}
\end{eqnarray}
Here, $v_{l,s}$ and $-v_{v}$ are  projections of the group velocities of the fundamental, Stokes and vibration waves on the z-axis,  $N$ is the number density of the vibrating molecules, $\alpha $ is the molecule polarizability,   $\tau_v$ is phonon lifetime.
Equations~(\ref{l})-(\ref{q}) are similar to those describing TWM of contra-propagating waves in NIMs for the phase-matching scheme of co-propagating fundamental and Stokes waves, where vibration wave counter-directed to the Stokes wave, which is depicted in Fig.~\ref{phf1}(a). In the case of continuous waves and neglected depletion of the fundamental wave,  resonance enhancement becomes possible, similar to those, depicted in Fig. \ref{f21} . On the contrary, only standard exponential behavior is possible in the case of Fig.~\ref{phf1}(b). Corresponding transmission factors in the vicinity of first ``geometrical'' resonance are shown in Fig.~\ref{phf2}.
In the resonance, $T_s^{\upuparrows}\rightarrow\infty$, which is due to the approximation of constant control field. Conversion of the control field to the Stokes one and excited molecule vibrations would lead to saturation of the control field which limits the maximum achievable amplification. Strong amplification in the maximums indicates the possibility of self-oscillations and thus creation of \emph{mirrorless} optical parametrical oscillator with unparalleled properties.
Figures~\ref{phf2}(a,b) indicate the possibility to \emph{fit in} the effective conversion length within the crystal of a given thickness and to significantly \emph{concentrate} the generated Stokes field nearby its output facet. Such atypical extraordinary behavior in readily available crystals may find exciting applications. However, the estimates have shown that intensity of the fundamental field which is to attain such extraordinary  amplification appears close to the optical breakdown threshold \cite{ph}. It is because of fast phonon damping and corresponding high rate of energy conversion  of the fundamental beam in heat.

Below, we show that such seemingly unavoidable obstacles can be overcome in short pulse regime.
For the sake of simplicity, we consider model of a rectangular pulse of input fundamental radiation with the pulse duration  $\tau_p\ll \tau_v$. A  seeding Stokes wave is assumed a weak continuous wave.
In the  moving coordinate frame associated with this pulse and within its range, complex amplitudes of  two other
interacting fields become time independent. Then the  analytical solution  to the equations can be found for amplification corresponding to negligible depletion of the pump due to the conversion. Numerical solution is found  for the opposite case. Inside the crystal, the boundary conditions must be fulfilled not at the boundaries of the medium but at the boundaries of the fundamental pulse. The latter is correct for the period of time after the instant when the generated Stokes and vibration pulses reach  the boundaries of the fundamental pulse.
Such approximation becomes true after travailing  a distance $l>L^{\max } $, where
$L^{\max } =\max \{L^s, L^v \}$,
$L^{s} ={l_p v_l}/{\left| v_s- v_{l}\right|}$ and
$L^{s,v} ={l_p v_l}/{\left( v_l +v_{v} \right)}$,  $l_p =\tau _p v_l$ is length of the fundamental pulse.
Hereinafter, the waves are referred to  as co-propagating  if the Poynting vector of the Stokes wave is co-directed with that of fundamental wave [Fig.~\ref{phf1}(a)], and as counter-propagating in the opposite case [Fig.~\ref{phf1}(b)]. Note, that  $v_{s}$ is negative for contra-propagating Stokes wave.

In the approximation of constant pump amplitude, in the coordinate frame  locked to the pump pulse, and within the pulse, equations for the generated Stokes and backward vibration waves take the form:
\begin{eqnarray}
\label{eq3}
 d\,Q/d\xi =-ig_v \mathcal{E} _s^* + Q K_v/l_v ,\quad
d\,\mathcal{E}_s /d\xi=ig_s Q^*.
\end{eqnarray}
Here, $\xi = z - v_l^{gr}t$, $l_v=\tau_v v_v$ is the phonon mean free path, $g_v= K_v N(d\alpha/dQ)\mathcal{E} _l /( 4\omega _v v_v)$, $K_{v}=v_{v}/(v_{l}+v_v)$,
$g_s=K_s N(d\alpha/dQ)\mathcal{E}_l \pi \omega_s^2/( k_s c^2)$, $K_{s}=v_{s}/(v_{s}-v_l )$; $v_s, k_s>0$  for co-propagating and $v_s, k_s<0$ for counter-propagating beams. Since the Stokes frequency is less than that of the fundamental one,  $v_s>v_l$  and $v_l\gg v_v$.

Equations (\ref{eq3}) are similar to those describing  CW TWM,  \cite{ph} except the boundary conditions. They correctly describe possible huge amplification of the  Stokes signal  until  relatively small part of the strong input laser beam is converted.
For co-directed laser and Stokes waves, the boundary conditions  are:
\begin{eqnarray}
\label{eq4}
\mathcal{E} _s (\xi =0)=\mathcal{E} _s^0, \quad 
 Q(\xi =l_p )=0.
\end{eqnarray}
In the opposite case,  they are written as
\begin{eqnarray}
\label{eq5}
 \mathcal{E} _s (\xi =l_p )=\mathcal{E} _s^{l_p }, \quad
 Q(\xi =l_p )=0.
\end{eqnarray}

The analysis of solution to Eqs.~(\ref{eq3}) shows that,  in the given approximation of neglected depletion of the fundamental wave, amplification of \emph{co-directed} signal  tends to infinity, when  the pulse energy approaches the \emph{resonance} value corresponding to  $g l_p = \pi/2$, where $g=\sqrt {g_v^* g_s }$.  This indicates the possibility of \emph{huge enhancement} of the conversion efficiency.
Respective intensity of the fundamental field $I_{\min }^p$ is given by the equation
\begin{equation}
\label{eq11}
I_{\min }^p =\frac{K_v}{K_s }\frac {cn_s \lambda _{s0} \omega _v}
{16\pi ^3v_v \tau_v ^2}\left| N\frac{\partial \alpha}{\partial Q}\right|^{-2},
\end{equation}
where, $n_s$ is refractive index at $\omega _v$ and $\lambda_{s0}$ is the wavelength in vacuum.
From comparison with the corresponding value $I_{\min}$ for the CW regime \cite{ph},  one concludes
\begin{equation}
\label{eq12}
\frac{I_{\min }^p }{I_{\min } }=\frac{K_v}{K_s }\approx \frac{v_v }{v_l }\frac{v_s -v_l }{v_s }.
\end{equation}
For the crystal parameters, which are characteristic for calcite \cite{Alf} and diamond \cite{G,An,Chen},  Eq.~(\ref{eq12}) yields $I_{\min }^p /I_{\min } \approx 10^{-11}$ and, hence, suggests a decreases of $I_{\min }^p$ down to $I_{\min }^p \sim 10^7$~W/cm$^2$. The latter is achievable with commercial femtosecond lasers and falls below the optical breakdown threshold for most transparent crystals.

Equation~(\ref{eq12}) displays  two factors that determine
substantial decrease of $I_{\min}^p$ in pulsed regime compared to that in CW one. First factor is the ratio of group velocity of the elastic wave to that of the fundamental one, $v_v/v_l $, which is on the order of $\sim 10^{-8}$. This factor is attributed to the fact that phonons generated on the front edge of the laser pulse propagate in the opposite direction and, hence, exit very fast, practically with the optical group velocity, from the fundamental pulse zone before it is dissipated.  Hence, effective phonon mean free pass becomes commensurable with the fundamental pulse length.  This mitigates the detrimental effect of phonon damping. With increase of $v_v$, phonon mean free path grows, which decreases both  $I_{\min }^p$ and  $I_{\min }$ in a way that the advantage of pulse regime over CW regime diminishes.
The second factor in Eq.~(\ref{eq12}) determines  further decrease of $I_{\min }^p $  due to small optical dispersion in the transparency region of the crystals. The fact that the Stokes pulse surpasses  the fundamental one slowly  increases   significantly the effective NLO coupling length.

To investigate the regimes of significant energy conversion, a set of partial differential Eqs.~(\ref{l})-(\ref{q}) was solved numerically in three steps: TWM in the vicinity of  the entrance, inside and  in the vicinity of  the exit from the Raman slab. Simulations for the first and third  intervals were made in the laboratory reference frame with the boundary conditions applied to the corresponding edges of the slab.  The propagation process inside the slab was simulated in the moving frame of reference  with the boundary conditions applied to the pulse edges. Such an approach  allowed for significant reduction of the computation time because, for each given instant, the integration was required only through the space interval inside the fundamental pulse and not through the entire medium.
 Shape of the fundamental pulse was chosen nearly rectangular and symmetric with respect to its center
$$\mathcal{E}_l =\frac{1}{2}\mathcal{E}_l^0 \{\tanh[( t_0+t_p-t)/t_f]-\tanh[(t_0-t)/t_f]\}.$$
The slope  $t_f =0.1$, the pulse duration at half-maximum $t_p=1$ and its delay $t_0 =0.6$ were scaled  to the fundamental pulse width $\tau_p$. The amplitude of input CW Stokes signal was chosen $\mathcal{E}_s^0=10^{-5}\mathcal{E}_l^0$.
Numerical investigations were done for the model with  parameters typical for  calcite \cite{Alf} ($\omega_v= 1086$~cm$^{-1}$) and diamond \cite{G,An,Chen} ($\omega_v= 1332$~cm$^{-1}$): carrying wavelength  $\lambda_l=800$~nm, pulse duration  $\tau_p=60$~fs, $\omega_v= 1332$~cm$^{-1}$,   $\tau_v= 7 ps$, $v_l=1.228\cdot10^{10}$~cm/s, $v_s=1.234\cdot10^{10}$~cm/s, $v_v=100$~cm/s for co-propagating and  $v_v=2000$~cm/s for counter-propagating waves, $Nd\alpha/dQ=3.78\cdot10^7$~(g/cm)$^{1/2}$, the crystal length L=1cm.

\begin{figure}[!t]
\begin{center}
\includegraphics[width=.45\columnwidth]{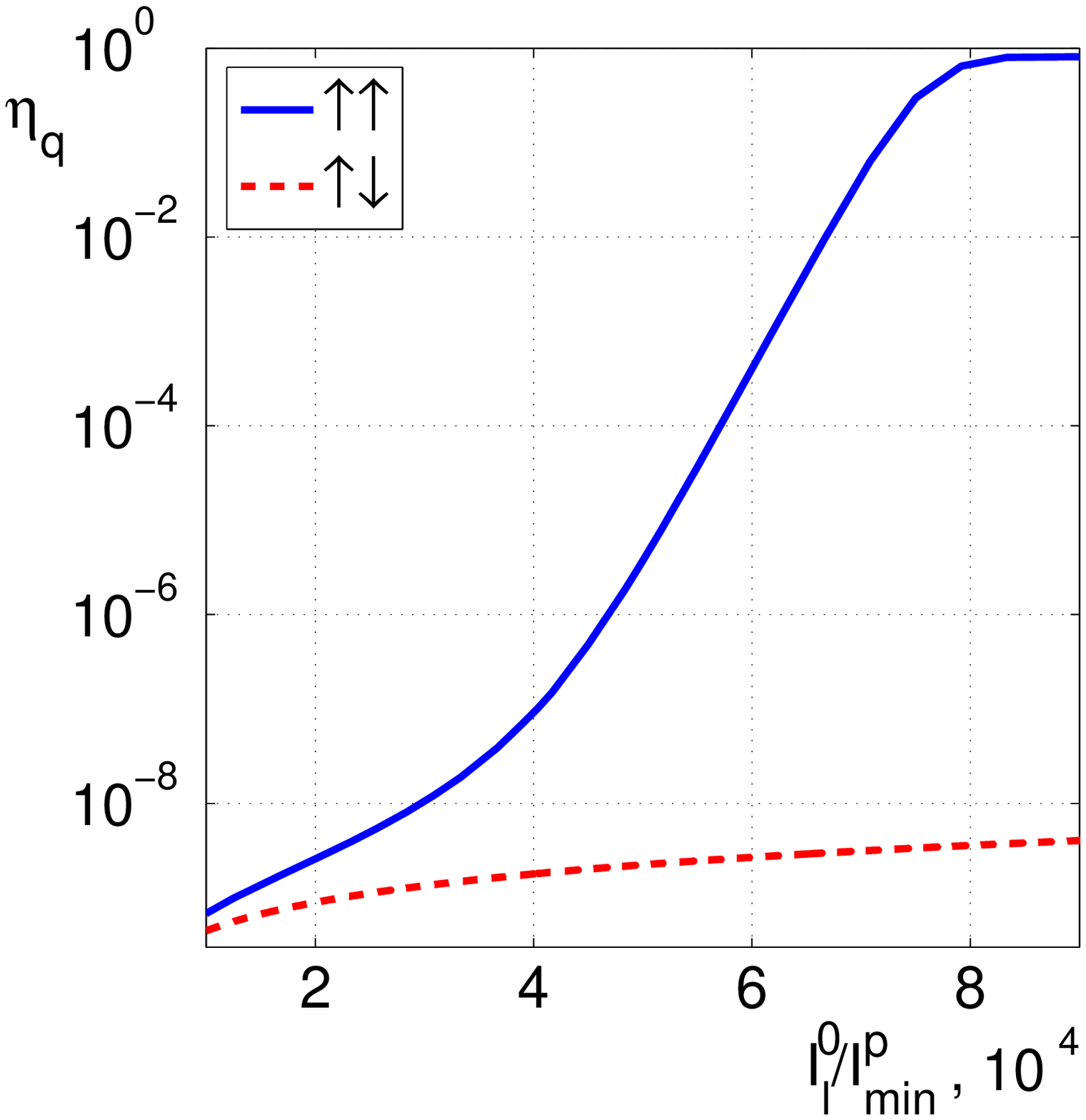}
\includegraphics[width=.53\columnwidth]{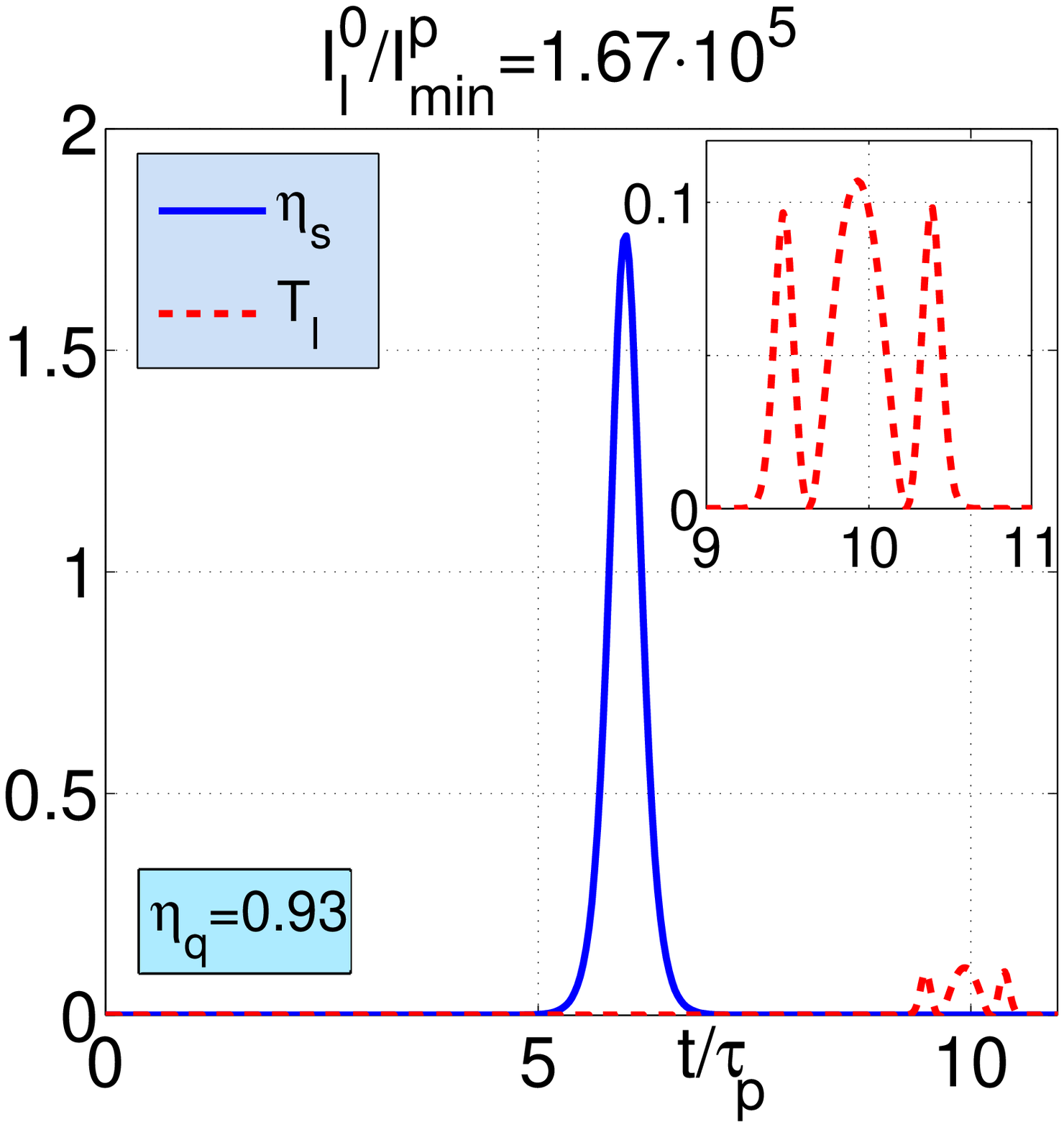}\\
(a)\hspace{40mm} (b)
\caption{\label{f5} a. Conversion efficiency vs. intensity of the input  pump for co-propagating (solid line)ss and contra-propagating  (dash line) geometries. b. Output Stokes (solid line) and fundamental (dash line) pulses  for co-propagating coupling. }
\end{center}
\end{figure}

Figure \ref{f5}(a) displays  the output quantum conversion efficiency
 $\eta_{q}=(\omega_l/\omega_s)\cdot\int_tI_s(z,t)dt/\int_tI_l(z=0,t)dt$ vs. input pulse intensity, both for co-propagating  ($z=L$) and counter-propagating ($z=0$) geometries.  Many orders increase of the conversion efficiency due to BW effect in the case of co-propagating waves is explicitly seen. Saturation at $I_l^0/I_{min}^p > 7\cdot10^4 $ is due to depletion of fundamental radiation caused by conversion to Stokes radiation. Figure \ref{f5}(b) displays a shortened Stokes output pulse that has surpassed the inhomogeneously depleted and broadened fundamental pulse (zoomed in the inset).

\begin{figure}[h!]
\begin{center}
\includegraphics[width=.55\columnwidth]{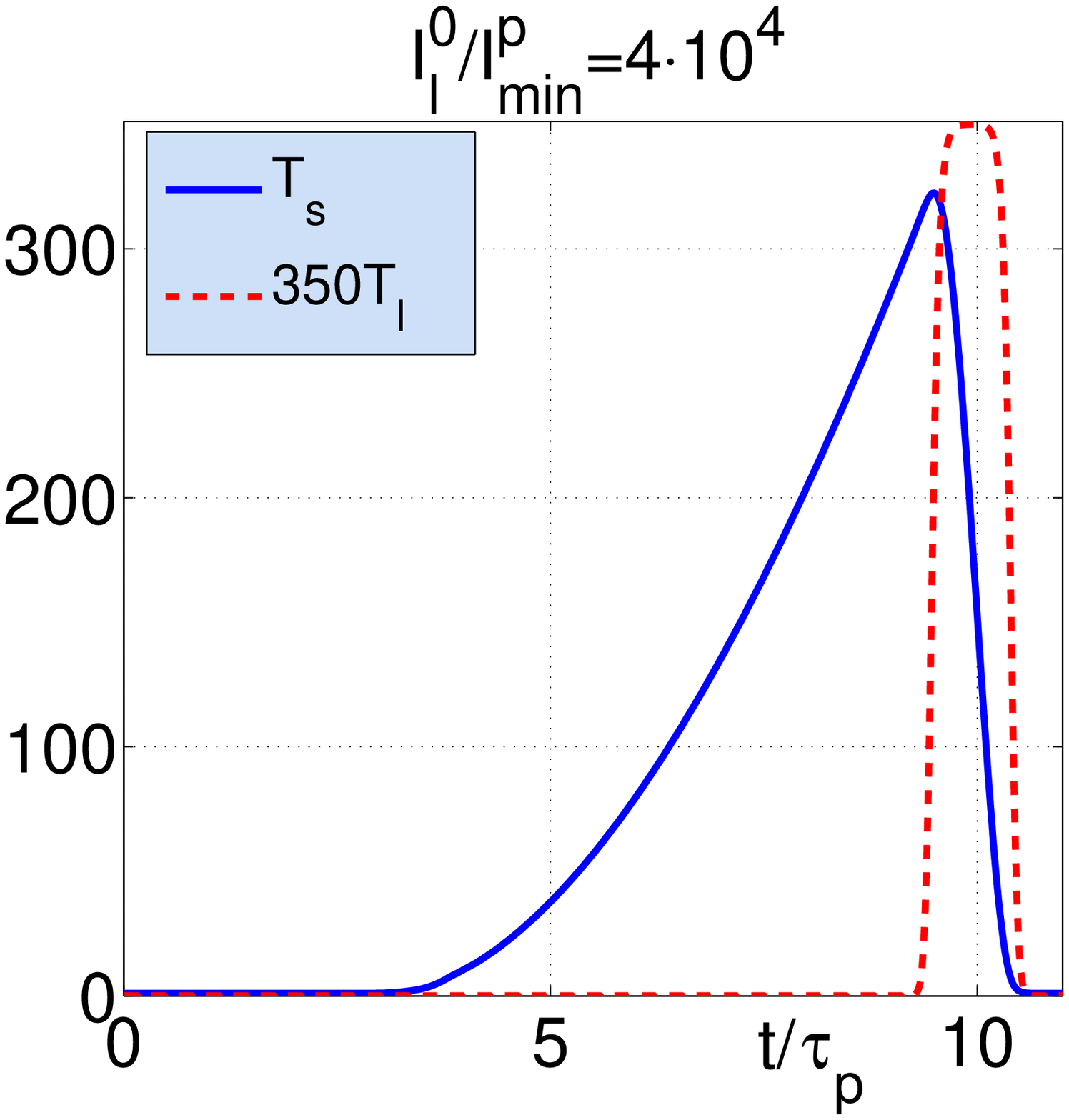}
\includegraphics[width=.43\columnwidth]{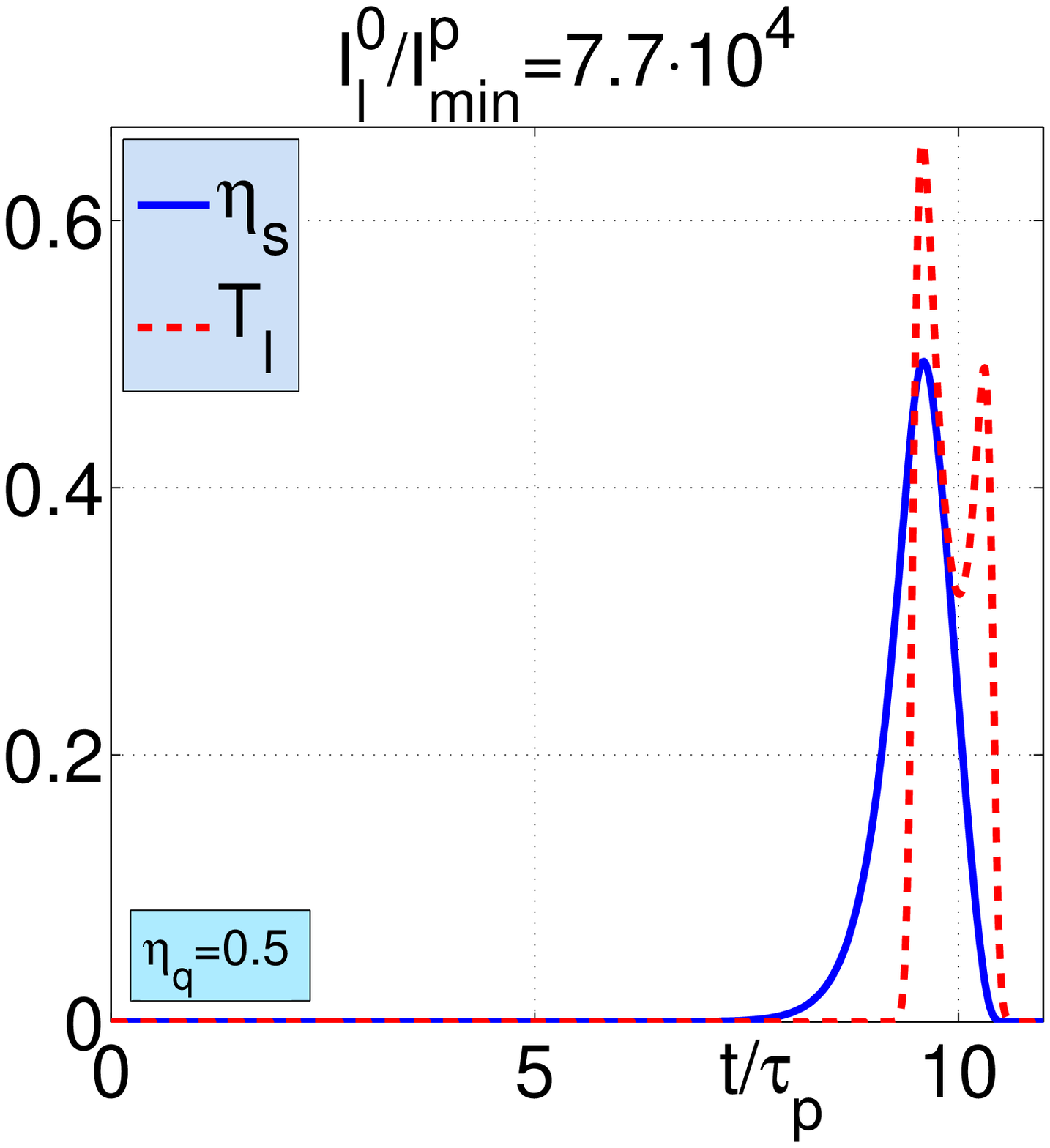}\\
(a)\hspace{40mm} (b)\\
\includegraphics[width=.45\columnwidth]{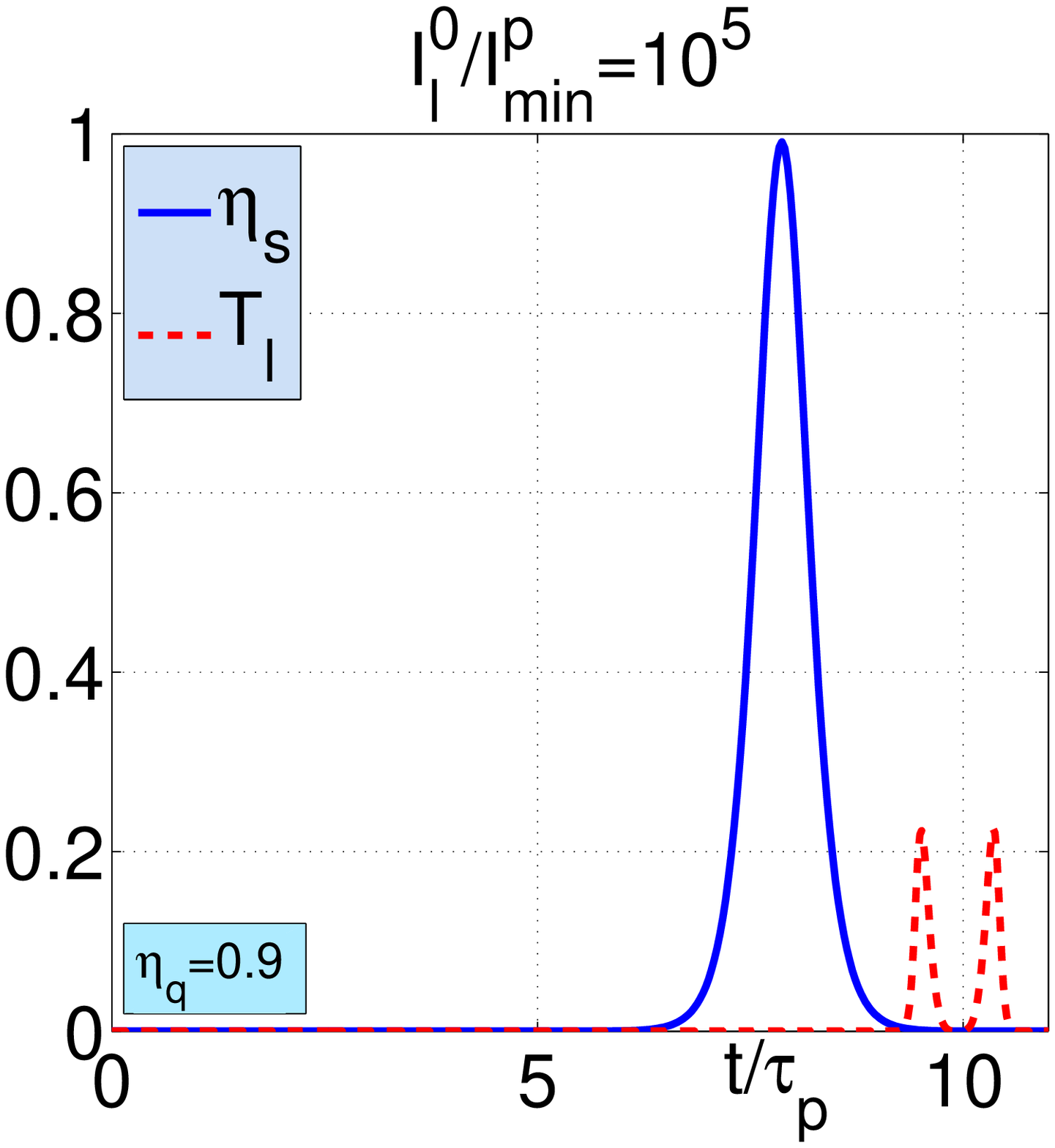}
\includegraphics[width=.43\columnwidth]{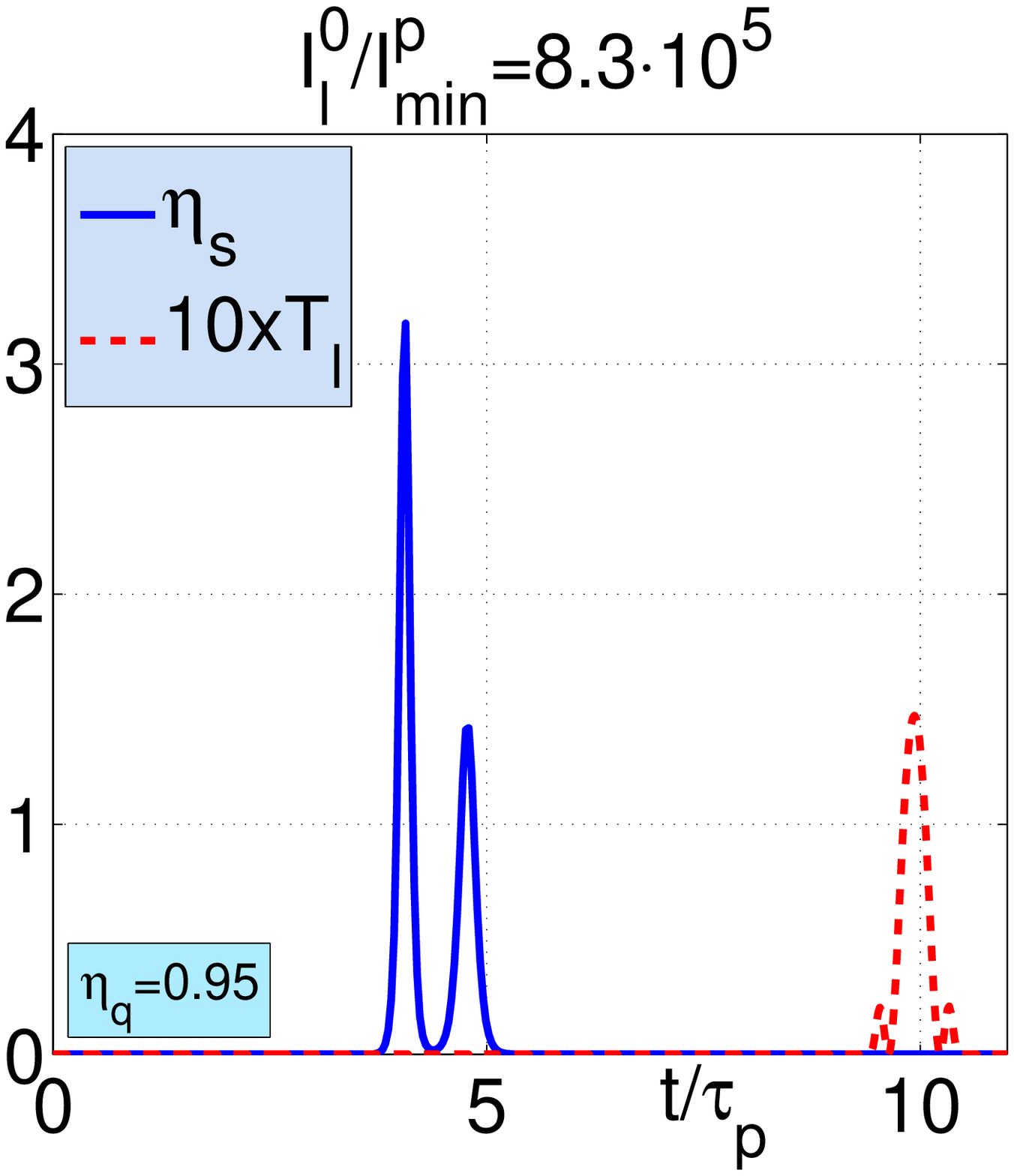}\\
(c)\hspace{40mm}(d)
\caption{\label{f8}  Changes in the shapes of  generated Stokes  (solid line) and transmitted fundamental  (dash line) output co-propagating pulses with  the increase of energy of the input fundamental pulse.  $\eta_q$ is corresponding  conversion efficiency.
}
\end{center}
\end{figure}

The simulations also show that shapes of the output Stokes and fundamental pulses  differ and vary significantly depending on the intensity of the input fundamental wave.  Figure~\ref{f8}(a) depicts  amplified output pulse  $ T_s =\left| {{\mathcal{E} _s (L,t)}/{\mathcal{E} _s^0 }} \right|^2$ of Stokes radiation for relatively small  depletion due to conversion  of the fundamental beam. Here, shape of the fundamental pulse is unchanged.  The shape of the amplified Stokes pulse is different and determined by the fact that $v_s>v_l$ and Stokes pulse has surpassed the fundamental one.
In contrast, the quantum conversion  becomes significant  in Figs.~\ref{f8}(b)-(d).  Corresponding depletion of the output fundamental pulse  and changes in its shape are explicitly seen.
Note that in the case depicted in Fig.~\ref{f8}(d), the output Stokes pulse significantly overtakes the pump pulse. In the latter case, major  conversion occurred inside, far from the crystal edges, and then both pulses propagated  without interaction. It is seen that output Stokes pulse narrows with the increase of energy of the fundamental pulse.  Here, crystal length of 1cm fits L/l$_p$ = 1357 input pulse lengths. The threshold intensity  $I^p_{min} = 6\cdot 10^6$ W/cm$^2$ corresponds to 60 fs pulse of about 5 $\mu$J energy focused to the spot of diameter D = 100 $\mu$m. Intensity of the seeding Stokes signal was chosen   $I^0_s/I^0_l = 10^{-10}$.

Note, that the described NLO propagation process  is in a striking contrast with their  positive group velocity counterparts \cite{ShB,Boy}, including  acoustic waves with  energy fluxes  directed against  that of EM waves \cite{Bob}.  The proposed here concept is different from the one earlier proposed in \cite{Har} and does not require periodic poling of quadratic nonlinear susceptibility of crystals at the nanoscale  as described in \cite{Kh} (and in the references therein).

\section{Conclusion}
The possibility of creation of remotely all-optical controlled frequency up-converting and amplifying nonlinear optical sensor (image and data processing chips) is investigated and proved with numerical simulations.
Novel photonic materials are proposed which support coexistence of ordinary and backward waves that can be phase matched. Such nonlinear-optical materials  enable greatly enhanced nonlinear-optical frequency-conversion processes that can be employed for creation of optical sensors with unparallel operational properties.
Two different classes of such materials are proposed: metamaterials with specially engineered spatial dispersion and crystals that support optical phonons with negative phase velocity. Unlike current mainstream in fabricating negative index metamaterials, both options do not rely on nanoresonators that provide negative optical magnetism and thus constitute current mainstream in fabricating negative index metamaterials. Extraordinary properties of coherent frequency conversion processes in the proposed materials  are numerically simulated both in continuous wave and in short pulse regimes.

As an example of dispersion engineering, a metamaterial made of standing carbon nanotubes is described which supports coexistence of ordinary fundamental and backward second harmonic electromagnetic waves that can be phase matched. Extraordinary properties of such frequency-doubling metamirror operating in short-pulse regime appear in striking contrasts with second harmonic generation in ordinary materials.

Fabrication of specially shaped nanostructures which enable negative optical magnetism is challenging task that relies on sophisticated methods of nanotechnology.  Engineering  a  strong  fast quadratic  NLO response by such mesoatoms also presents a  challenging goal not yet achieved. This work proposes a different paradigm, which is to mimic similar extraordinary backward-wave frequency conversion processes through  making use of readily available Raman active crystals.
Basic underpinning idea is to replace one of the coupled  backward optical electromagnetic mode,  which existence is commonly attributed to negative-index metamaterials,  by optical phonons - the elastic wave with  negative group velocity. Operation in short pulse regime is proposed to remove  a severe detrimental factor imposed by  fast phonon damping. Significant decrease of the required minimum intensity of fundamental radiation, as compared with that in the continuous-wave regime, is predicted down to that provided by commercial lasers. Unparallel properties of the proposed short-pulse process are  numerically simulated and the possibility of huge enhancement of quantum conversion efficiency as well as of tailoring duration and shapes of the  generated and transmitted fundamental pulses  are predicted.

Further elaboration of the proposed concept  allows  to utilize  the revealed extraordinary features for  creation of a family of unique photonic devices with advanced functional properties  such as  nonlinear-optical mirrors with frequency-dependent controllable reflectivity,   unidirectional optical amplifiers, frequency narrow band filters, switches and cavity-free optical parametric oscillators, which can be exploited for optical sensing.
There exist many dispersive materials and readily available Raman-active transparent crystals that support electromagnetic and elastic waves with negative group velocity which can be utilized for the given purpose. Some of them  may offer the opportunities for further optimization of  unparallel operational characteristics of the proposed photonic devices.

\section{Acknowledgments}
 This work was supported in parts by the Air Force Office of Scientific Research
(Contract  No FA950-12-1-298), by the  National Science Foundation (Grant No ECCS-1028353), by the
Academy of Finland and Nokia through the Center-of-
Excellence program, by the Presidium of the Russian
Academy of Sciences (Grant No 24-31), and by the
Russian Federal Program on Science, Education and In-
novation ( Grant No 14.A18.21.1942).


\end{document}